\documentclass[onecolumn]{els-mrw} 
\usepackage{amsmath,amssymb,amsfonts,amsthm,makeidx,graphicx}
\usepackage{txfonts}
\usepackage{helvet}

\begin{document}
\chapter{Chaotic Dynamics and Quantum Transport}\label{chap1}
\author[1,2]{Andrey R. Kolovsky}%
\address[1]{\orgname{Siberian Branch of Russian Academy of Sciences}, \orgdiv{L. V. Kirensky Institute of Physics}, \orgaddress{660036 Krasnoyarsk, Russia}}
\address[2]{\orgname{Siberian Federal University}, \orgdiv{School of Engineering Physics and Radio Electrtronic }, \orgaddress{660041 Krasnoyarsk, Russia}}
\articletag{Chapter Article tagline: Sept. 1, 2025}
\maketitle

\begin{glossary}[Keywords]
Quantum Chaos, master equation, pseudoclassical approach, two-terminal transport, ballistic and diffusive regimes
\end{glossary}

\begin{abstract}[Abstract]
This chapter gives an overview of transport problems in which the system's chaotic dynamics play a crucial role. We begin with single-particle transport and then discuss conservative and dissipative systems of identical particles, following the historical way development of quantum-chaos theory over the past 40 years. We also include brief descriptions of key laboratory experiments on the discussed transport problems.
\end{abstract}

\section{Introduction}
The term `quantum transport' spans a wide variety of phenomena, including the wave-packet dynamics in the momentum or coordinate spaces, the particle transport in experiments where one measures the particle current (for example, electric current), the heat or energy transport, spin transport, and so on. It follows from this incomplete list that the field of quantum transport is too diverse to be reviewed in a single book, let alone a single chapter. Thus, to keep the chapter length to a reasonable limit one has to narrow the subject.  As the chapter title indicates, the main criterion for choosing topics to discuss is their relevance to Quantum Chaos \cite{ Gian91,Haak91,Stoc99}.  However, the criterion of relevance to chaos alone is not sufficient because there are many different formulations of the transport problem depending on the considered  physical system and the abilities of the laboratory experiment. We sort all transport problems into three groups according to the quantum-mechanical formalism, which is used to analyze the problem. The first group includes problems of single-particle quantum mechanics where knowledge of the single-particle wave-function is sufficient to explain the results of a laboratory experiment. The second group consist of many-body transport problems where inter-particle interaction is the key ingredient. Here the wave function obeys the many-body  Schr\"odinger equation and the analysis is usually done by using the second quantization formalism. Needless to say that one should distinguish between Bose and Fermi quantum statistics.  Finally, the largest group of transport problems concerns systems which admit a description in terms of the reduced density matrix.  This group includes, in particular, the so-called boundary-driven systems where the system of interest, for example, one-dimensional spin chain is coupled at its ends to thermal reservoirs \cite{Bert21,Land22}. In what follows we shall provide examples of quantum transport problems from each of the above-mentioned groups and build the discussion of the related problems around these examples. 

It should also be stated from the very beginning  what is not considered in this chapter, i.e., the scope boundaries. First of all, we do not review sophisticated theoretical approaches to quantum transport like the Green's functions methods, diagram techniques, etc. and do not address general aspects of the transport theory like fluctuation-dissipation theorem or universal conduction fluctuations. These aspects and methods of transport theory can be found, for example, in the books \cite{Datt95,Ditt98}. In this chapter we always stay on the elementary level and use straightforward (analytical or numerical) methods to solve the quantum equations of motion. Finally, when discussing  systems of identical particles, we restrict ourselves to the case of Bose particles. The main reason behind this choice is that bosonic systems have well-defined classical counterparts and, thus, we can appeal to classical mechanics to get some intuition.

\section{Single-particle transport problems}\label{sec2}
Typical formulation of the quantum chaotic-transport problem in the single-particle case is as follows. We have a classical Hamiltonian system showing chaotic transport, which means that the classical trajectory is sensitive to the initial condition (more rigorously, has a positive Lyapunov exponent) and exhibits dynamics resembling  a random walk. We quantize this system and ask the question of how quantum mechanics modifies the classical result. Since in the quantum realm the notion of particle trajectory is poorly defined, the comparison is done in terms of averaged quantities -- the quantum average   $A_{qu}(t)=\langle\psi(t) | \widehat{A}|\psi(t)\rangle$ and the classical average $A_{cl}(t)=\langle A \rangle$, where the angular brackets denote averaging over the appropriate ensemble of initial conditions.  
\footnote{This ensemble is determined by the initial state of the quantum system and will be discussed later on in some detail. In the example considered below, where the kicked rotor is initially in its ground state, this ensemble corresponds to $I(t=0)=0$ and a random $\theta(t=0)$ uniformly distributed in the interval $[0,2\pi)$.} 
For the first time this question was addressed  in the pioneering  work by Casati {\em et al.} \cite{Casa79} where the authors compared the classical and quantum dynamics of the kicked rotor,
\begin{align}\label{KR}
H=\frac{I^2}{2\mu} + V\cos\theta \sum_n \delta(t-nT) \;.
\end{align}
In the chaotic regime the classical kicked rotor is known to show the diffusive growth of the kinetic energy $I^2/2\mu$, i.e., the mean squared angular momentum grows linearly in time  $\langle I^2 \rangle \propto t$.  It was found, however, that the quantum kicked rotor,  $I\rightarrow\hat{I}=-i\hbar\partial/\partial\theta$, exhibits the linear growth only for a finite time and then the energy either saturates at some value (if the ratio $T\hbar/\mu$ is an irrational number) or asymptotically grows as $t^2$ (if $T\hbar/\mu$ is a rational number).  
\footnote{ For the detailed analysis of the kicked rotor and related problems see chapter \cite{chapter_16} in this volume.}
The energy saturation was explained a few years later by drawing an analogy with Anderson localization and was labelled {\em dynamical localization}. On the formal level it means that eigenstates of the Floquet operator (the evolution operator over the driving period $T$),
\begin{align}\label{Floquet}
\widehat{U}=\widehat{\exp}\left[-\frac{i}{\hbar}\int_0^T \widehat{H}(t){\rm d}t\right]  \;,
\end{align}
are localized functions in the (angular) momentum space in the case of irrational $T\hbar/\mu$ and extended functions if $T\hbar/\mu=r/q$. We emphasize that this localization-delocalization transition  is no less interesting than seeking signatures of classical chaos in the quantum case. We shall constantly meet this transition, which is governed by some commensurability condition on the system parameters, in other systems considered in this section.
\begin{figure}[t]
\centering
\includegraphics[width=.5\textwidth]{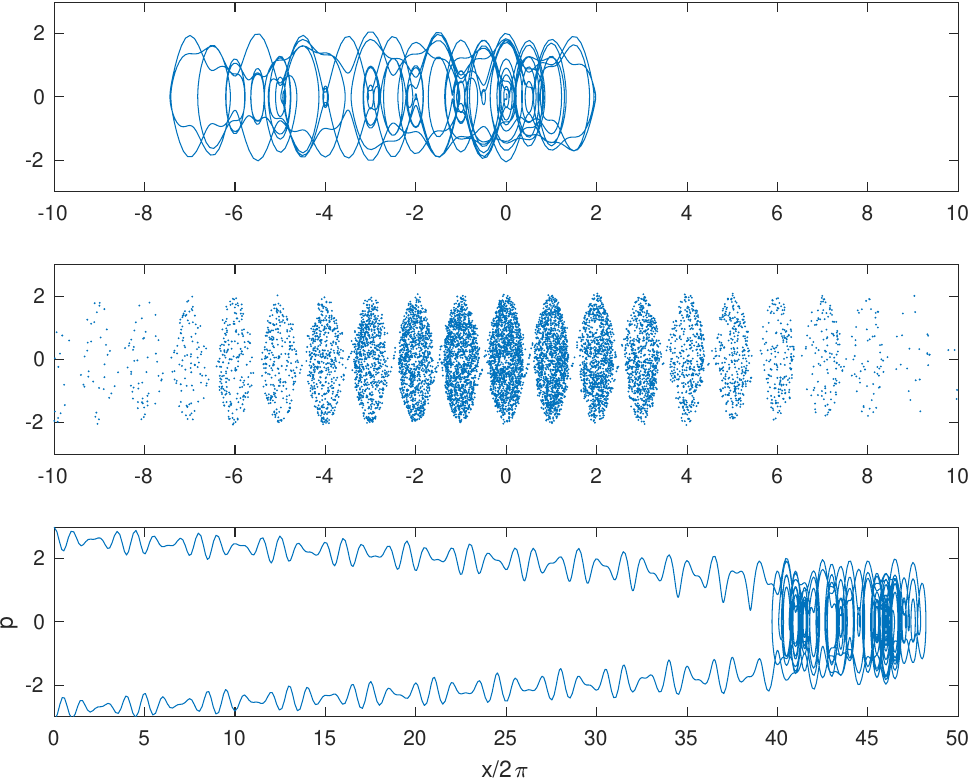}
\caption{(a) Typical chaotic trajectory in the phase space of the system (\ref{drm}). The system parameters are $m=1$, $V=-1$, $d=2\pi$, and $\omega=0.3$. Evolution time is 20 periods of the driving force. (b) Chaotic diffusion. Dots show final particle positions for 10000 initial conditions randomly distributed according to a narrow two-dimensional Gaussian centered at $x=p=0$. (c) Chaotic scattering in the case where the system Hamiltonian (\ref{drm}) is accompanied  by the Stark term $Fx$ with $F=0.01$. }
\label{fig2_1}
\end{figure}

\subsection{Quantum particle in a driven lattice}\label{sec2_1}
Another early example of chaotic quantum transport is a particle in the standing wave \cite{22},
\begin{align}\label{drm}
H=\frac{p^2}{2m} + V\cos(\omega t)\cos\left(\frac{2\pi x}{d}\right) \;.
\end{align}
For small values of  the driving frequency $\omega<d\sqrt{V/m}$ the phase portrait of the system (\ref{drm}) consists of the chaotic strip $-\infty<x<\infty$ and $|p|<p^*\sim\sqrt{Vm}$, where the classical particle moves randomly with $\langle x^2 \rangle \propto t$, and regular trajectories for $|p|>p^*$, where it moves ballistically. We focus on the chaotic motion, see Fig.~\ref{fig2_1}. The main question which was addressed in the cited work is the same as for the kicked rotor, i.e., quantum modification of classical chaotic diffusion. However, unlike the kicked rotor, the system (\ref{drm}) is translationally invariant, $H(x+d)=H(x)$, and, in this sense, is closer to transport problems of solid-state physics. In particular, the eigenstates of the Floquet operator (\ref{Floquet}), where $\widehat{H}$ is the quantum Hamiltonian  ($p\rightarrow\hat{p}=-i\hbar\partial/\partial x$) of the system (\ref{drm}), are now labeled by the quasimomentum $\kappa$ and, thus, its spectrum consists of the Bloch bands.
\footnote{Notice that for any system with the band spectrum $\langle x^2\rangle=\langle\psi(t) | x^2|\psi(t)\rangle$ asymptotically grows as $t^2$.}
It was shown in Ref.~\cite{22} that this spectrum can be decomposed into chaotic bands, which are associated with the chaotic component of the classical phase space, and regular bands, which are associated with the regular component. Yet, these two spectra are not completely independent. This is manifested by the avoided crossings between regular and chaotic bands. These avoided crossings have a pronounced effect  on the quantum transport leading to further acceleration of the classical diffusion. Nowadays, the physical effects due to `quantum mixing' of the regular and chaotic components of the classical phase space are known as the phenomenon of {\em chaos-assisted tunnelings}, see chapter \cite{chapter_17} in this volume. 

We also briefly discuss an abstract variant of the system (\ref{drm}) known as the kicked Harper \cite{Scha01},
\begin{align}\label{kicked_harper}
H=-\cos p + V\cos x \sum_n \delta(t-nT)  \;, \quad  |p|\le\pi \;, \quad |x|<\infty \;.
\end{align}
In the quantum case one defines the operator $\cos \hat{p}=\cos( -i\hbar \partial/\partial x)$  as a sum of two shift operators. For $V\gg 1$ the classical system (\ref{kicked_harper}) has no regular component which simplifies the analysis of the quasienergy spectrum. Excellent agreement of the statistics of energy gaps in the quasienergy spectrum with prediction of the Random Matrix Theory (RMT) 
\footnote{See chapter \cite{chapter_5} in this volume.}
was reported in Ref.~\cite{59}.  Also, it was shown in Ref.~\cite{59} that by adding to the Hamiltonian  (\ref{kicked_harper}) additional Stark term $Fx$  one can change the transport regime from the accelerated diffusion to the dynamical localization. 

The list of systems with chaotic transport in the classical case can be extended further but all these systems were mainly of academic interest, with little prospect of their experimental realization. This situation has cardinally changed after the tremendous progress in cooling a gas of neutral atoms to ultra-low temperatures, where atoms can be captured in the dipole potential of an optical lattice created by the standing laser wave. The first transport experiment with cold atoms in optical lattices reported the observation of Bloch oscillations \cite{Daha96}.  
\footnote{For the introductory review of {\em atomic} Bloch oscillations  we refer the reader to Ref.~\cite{63}.}
Due to the importance of Bloch oscillations for what follows we shall discuss this phenomenon in some detail in the next subsection.

\subsection{Bloch oscillations}\label{sec2_2}
Bloch oscillations are one of the most fascinating interference effects in single-particle quantum mechanics.  Its origin lies in the band spectrum of a quantum particle in a periodic potential, which leads to a counterintuitive response of the system to an external  static field/force. Considering for the moment only the ground Bloch band, the Hamiltonian of the quantum particle in a  tilted lattice, 
\begin{align}\label{drm_stark}
\widehat{H}=\frac{\hat{p}^2}{2m} + V\cos\left(\frac{2\pi x}{d}\right) +Fx \;,
\end{align}
can be approximated by the tight-binding Hamiltonian 
\begin{align} \label{ws}
\widehat{H}=\widehat{H}_0+ dF\sum_\ell  \ell |\ell\rangle\langle \ell | \;, \\
\label{tb}
\widehat{H}_0=E_0\sum_\ell |\ell\rangle\langle \ell |  - \frac{J}{2}\sum_\ell \left(|\ell+1\rangle\langle \ell | + h.c.\right) \;,
\end{align}
where $|\ell\rangle$ are the localized Wannier states,  
\footnote{Wannier states $|\ell\rangle\equiv\Phi_\ell(x)$ are found by integrating the Bloch states $\psi_\kappa(x)$ over the quasimomentum, $\Phi_\ell(x)=\oint \psi_\kappa(x) e^{i\ell\kappa}{\rm d}\kappa$. As a rule, the obtained function  $\Phi_\ell(x)$ are approximated by a Gaussian located at the $\ell$th well of the lattice.}
$J$ is the rate of inter-well under-barrier tunneling, and $d$ the lattice period. Diagonalizing the Hamiltonian (\ref{tb}) we obtain the cosine dispersion relation $E(\kappa)=E_0 - J\cos(d\kappa)$, which approximates the dispersion relation of the ground Bloch band, while diagonalizing the Hamiltonian (\ref{ws}) we obtain the so-called Wannier-Stark states,
\begin{align}\label{ws_states}
|\Psi_n\rangle=\sum_\ell {\cal J}_{\ell-n}\left(\frac{J}{dF}\right) |\ell \rangle 
\end{align}
(here ${\cal J}_{m}(z)$ is the Bessel function of the first kind) and the energy spectrum 
\begin{align}\label{ws_spectrum}
E_n=E_0+\hbar\omega_B n \;,\quad \omega_B=dF/\hbar \;.
\end{align}
The equidistant spectrum (\ref{ws_spectrum}) implies periodic oscillations of the quantum particle with an amplitude determined by  the localization length $L_{WS}\approx 2J/dF$ of Wannier-Stark states (\ref{ws_states}). 

As a side remark we mention that  one obtains the same result by using classical approach where the effective Hamiltonian of the lattice particle subject to a static force $F$ is given by
\begin{align} 
H_{eff}=E_0-J\cos\left(\frac{dp}{\hbar}\right) + Fx \;.
\end{align}
Indeed, solving the Hamiltonian equation of motion we have $p(t)\equiv\hbar\kappa(t)=Ft$, that is known as the Bloch acceleration theorem, and $x(t)=(Jd/\hbar\omega_B)\cos(\omega_B t)$. Thus,  the particle oscillates in the coordinate space with the Bloch frequency $\omega_B$, which is proportional to the static force magnitude, and the amplitude $x_{max}=Jd/\hbar\omega_B$, which is inverse  proportional to the static force. This approach,  where one puts into correspondence a quantum Hamiltonian in matrix form with an effective classical system,  is widely used in the theory of quantum chaotic transport.  We shall use this approach later on in Sec.~\ref{sec3}.  

It should be stressed that the used single-band approximation, which leads to the tight-binding Hamiltonian (\ref{ws}), is valid only for a small $F<\Delta/d$ where $\Delta$ is the energy gap separating the ground Bloch band from the rest of the energy spectrum. The latter condition ensures negligible inter-band Landau-Zener tunneling. If this is not the case, the problem should be analyzed in terms of the metastable Wannier-Stark states whose eigen-energies are complex numbers ${\cal E}_n=E_0+\hbar\omega_Bn-i\Gamma/2$ \cite{53}. In what follows we shall refer to the parameter region where the original quantum system can be described by its tight-binding version as the deep quantum regime.


\subsection{Driven tilted lattices}\label{sec2_3}
An example of a driven lattice was given in Eq.~(\ref{drm}).  We mention that  by using non-stationary optical lattices one can realize practically any kind of driving, including the DC-AC driving,
\begin{align} \label{ws_chaos}
\widehat{H}=\frac{\hat{p}^2}{2m} + V\cos\left(\frac{2\pi x}{d}\right) +Fx +F_\omega\sin(\omega t)x \;,
\end{align}
which is of particular interest due to its relation to the problem of crystalline electrons subject to static and alternating  electric fields. The system (\ref{ws_chaos}) was analyzed in full detail in the review \cite{53}, from which we borrow a few results.

In the deep quantum regime, where the Hamiltonian (\ref{ws_chaos}) is approximated by the tight-binding Hamiltonian
\begin{align} \label{ws_driven}
\widehat{H}=\widehat{H}_0+ dF\sum_\ell  \ell |\ell\rangle\langle \ell | + dF_\omega\sin(\omega t)\sum_\ell  \ell |\ell\rangle\langle \ell | \;,
\end{align}
the AC driving leads to the {\em induced tunneling}. 
\footnote{An extensive list of references on early studies of the system (\ref{ws_driven}) is given in  section 5.1 of  the review \cite{53}.}
The essence of this effect is as follows. As shown in the previous subsection, the static force $F$ leads to the ladder of localized Wannier-Stark states with the energy step determined by the Bloch frequency $\omega_B$. However, if the driving frequency  $\omega$ exactly matches the Bloch frequency, the periodic driving will couple all these localized  states into extended Bloch states and, thus, the quantum particle can again freely move in the lattice. The dispersion relation of these new Bloch states is given by the equation 
\begin{align} \label{ws_dispersion}
E(\kappa)=J{\cal J}_1\left(\frac{F_\omega}{F}\right)\cos(d\kappa+\phi)   \;,
\end{align}
where we include the phase $\phi$ for generality. This phase $\phi$ is the phase difference between the Bloch-oscillation and driving-field phases and it can be well controlled in laboratory experiments with cold atoms \cite{Hall10}.  

Eq.~(\ref{ws_dispersion}) can be generalized to the case where the ratio between the Bloch and driving frequencies is a rational number, $\omega/\omega_B=r/q$. Then the quasi-energy band (\ref{ws_dispersion}) splits into $r$ subbands and the widths of these subbands progressively decrease as either $r$ or $q$ are increased. For irrational $\omega/\omega_B$ the spectrum becomes pure point which implies localized eigenstates. This localization-delocalization transition is similar to the Aubry-Andr\'{e} localization  in one-dimensional quasi-periodic lattices \cite{Aubr80}.  We shall come back to this effect in the subsection \ref{sec2_5} in relation to eigenstates  of quantum particle in tilted two-dimensional lattices.
 
\subsection{Chaotic bands}\label{sec2_4}
The  phenomenon of induced tunneling just discussed above refers to the deep quantum regime of cold atoms in shallow lattices and has nothing to do with chaotic transport. To have chaotic dynamics one has to use a deep optical lattice and strong driving, that brings the quantum system (\ref{ws_chaos}) closer to  its classical counterpart. Notice that strong driving unavoidably violates the single-band approximation and, hence, we need computational support to understand the quantum dynamics of the driven system (\ref{ws_chaos}). 
\begin{figure}[t]
\centering
\includegraphics[width=.55\textwidth]{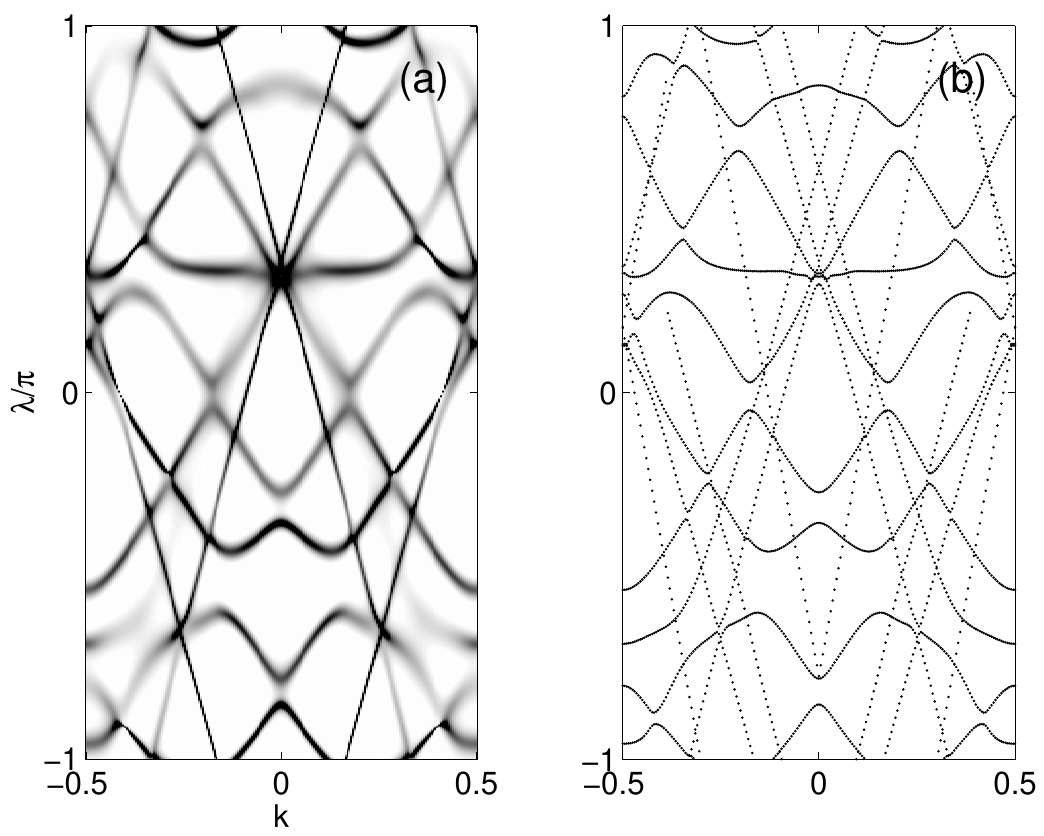}
\caption{(a) Wigner delay time as a grayscale map. (b )The real part of the quasienergy spectrum where we plot the first 13 most stable states for each $\kappa$. $\lambda$, $-\pi\le \lambda <\pi$, is the quasienergy in units of  $\hbar'\omega$. The system parameters are $\omega=10/6$, $\omega/\omega_B=1$, $\varepsilon=F_\omega/\omega^2=1.5$, and the scaled Planck constant $\hbar'=0.5$. The figure is borrowed from Ref.~\cite{53}.}
\label{fig2_4}
\end{figure}

It is convenient to use the canonical substitution $p\rightarrow p-(F_\omega/\omega)\cos(\omega t)$ and $x\rightarrow x+(F_\omega/\omega^2)\sin(\omega t)$, which transforms the system Hamiltonian to the form 
\begin{align} \label{ws_chaos_2}
\widehat{H}=\frac{\hat{p}^2}{2} + \cos [x+\varepsilon\sin(\omega t)] +Fx  \;,\quad \varepsilon=\frac{F_\omega}{\omega^2} \;.
\end{align}
(From now on we use the scaled variables where the parameter characterizing the degree of classicality is the scaled Planck constant $\hbar'=  2\pi\hbar/\sqrt{md^2V}$ entering the definition of the momentum operator.) For $F=0$ the phase portrait of  the classical counterpart of the system (\ref{ws_chaos_2}) is similar to the system (\ref{drm}), i.e., it consists of a chaotic strip for  $|p|<p^*$, where the particle randomly walks along the lattice,  and regular component for  $|p|>p^*$, where the particle moves ballistically. 
The case $F\ne0$ is more subtle. Here the problem is formulated in terms of chaotic scattering. Namely, the particle comes from minus infinity and enters the chaotic strip near the turning point of the linear potential $Fx$. There it walks randomly along the lattice for some time and then escapes back to minus infinity, see Fig.~\ref{fig2_1}(c). The time $\tau$, which the particle spends in the chaotic region, is know as the dwell or {\em delay time}.  For the classical chaotic scattering the delay time is distributed according to the universal exponential law
\begin{align} \label{cl_scattering}
P_{cl}(\tau)=\frac{1}{\bar{\tau}}\exp\left(-\frac{\tau}{\bar{\tau}}\right) \;,
\end{align}
where $\bar{\tau}$ is the mean delay time which is the system specific.

In the quantum case the main object to study is eigenvalues of the Floquet operator (\ref{Floquet}) which, as was already mentioned, are labeled by the quasimomentum $\kappa$. In other words, the spectrum of the Floquet operator consists of quasienergy bands  which in the case $F\ne0$ are complex numbers, ${\cal E}_\alpha(\kappa)=E_\alpha(\kappa)-i \Gamma_\alpha(\kappa)/2$ (here $\alpha$ is the band index).   We notice  that instead of the widths $\Gamma_\alpha(\kappa)$ one can consider the Wigner delay time $\tau(E,\kappa)=-i({\rm d} r/{\rm d}E)/r$ where $r=e^{i\theta(E)}$ is the reflection amplitude. (Notice that for tilted lattices reflection probability $|r|^2$ is always unity.)  The typical example of the quasienergy spectrum of the Floquet operator is depicted in Fig.~\ref{fig2_4}, where the right panel shows the real part $E_\alpha(\kappa)$ of the quasienergy  spectrum and the left panel the Wigner delay time $\tau(E,\kappa)$ which carries information on the imaginary part of the spectrum $\Gamma_\alpha(\kappa)$.  

The fact that the system (\ref{ws_chaos_2}) is chaotic in the classical limit implies the system is also chaotic in the sense of quantum chaos, i.e., that its spectral statistics follows the predictions of random-matrix theory.  
\footnote{Random-matrix theory for chaotic wave scattering is discussed in chapter \cite{chapter_4} in this volume. }
The statistics of the Wigner delay for different ratios $\omega/\omega_B=r/q$ were analyzed in a series of works by Gl\"uck {et al.}, collected in the review \cite{53}. In the same works it was argued there that   the denominator $q$ in the commensurability condition $\omega/\omega_B=r/q$ has the meaning of the number of open channels in the theory of quantum scattering. For quantum chaotic scattering with $q$ open channels RMT predicts the following distribution for the scaled (in units of the Heisenberg time) Wigner delay time,
\begin{align} \label{RMT_scattering}
P_{qu}(\tau)=\frac{1}{q!}\frac{1}{\tau^{q+2}}\exp\left(-\frac{1}{\tau}\right) \;.
\end{align}
Comparison of the Wigner delay time statistics for the system (\ref{ws_chaos}) with Eq.~(\ref{RMT_scattering})  revealed  excellent agreement \cite{53}. We would like to draw the reader's attention to the fact that the quantum distribution of the delay time $\tau$, which is the quantum analogue of the the classical delay time, fundamentally differs from Eq.~(\ref{cl_scattering}) and depends on the commensurability condition $\omega/\omega_B=r/q$. This condition is a clear manifestation of quantum interference which strongly modifies classical dynamics. The importance of this commensurability condition  is further highlighted in Sec.~\ref{sec2_5}.  

\subsection{Quantum billiards}\label{sec2_x}
It is the appropriate place here to mention other systems with chaotic scattering -- the quantum billiards and quantum graphs, see chapters \cite{chapter_12,chapter_13,chapter_14,chapter_21} in this volume. Strictly speaking, the transport problem for these systems  falls in the category of many-body boundary-driven systems (see Sec.~\ref{sec4} below), but in the case of vanishing inter-particle interaction, it can be reformulated as a problem of single-particle quantum mechanics. A typical example is the electron transport through the semiconductor quantum dot of a given shape in the free-electron-gas approximation \cite{x}. Within this approximation, the main object of study is the transmission probability $|t(E)|^2$ for the electron de Broglie wave, which enters the Landauer-B\"uttiker formula for the system conductance \cite{Datt95}. If the quantum dot has the shape of a chaotic billiards, the transmission probability is known to be a rather complicated function of the electron energy $E$. To analyze this function, one usually employs the semi-classical approach where the starting point is the set of classical trajectories connecting the incoming and outgoing channels (i.e., two contacts which are attached to the quantum dot), see chapter \cite{chapter_3} in this volume. Similar systems are a quasi-2D microwave resonators (microwave billiards)  \cite{Stoc99} and a net of coaxial cables \cite{kottos}. In these systems $|t(E)|^2$ is transmission probability for the electro-magnetic wave. 

\subsection{Landau-Stark states}\label{sec2_5}
We come back to the lattice systems. In this subsection we consider a quantum particle (to be certain, an electron) in a two-dimensional lattice in the presence of an in-plane electric field ${\bf F}$ and a normal to the lattice plane magnetic field $B$. The quantum Hamiltonian of this system reads 
\begin{align} \label{2d}
\widehat{H}=\frac{1}{2m}\left[\hat{{\bf p} } - \frac{e}{c}{\bf A}({\bf r})\right]^2 + V({\bf r}) +e{\bf F\cdot r}  \;, \quad 
V({\bf r}+{\bf R}_j)=V({\bf r}) \;,
\end{align}
where ${\bf R}_j$ are the primitive vectors of the lattice and the standard choice for the vector potential ${\bf A}({\bf r})$ is either the Landau gauge ${\bf A}=B(0,x)$ or the symmetric gauge ${\bf A}=B(-y/2,x/2)$. As concerns the lattice, one should distinguish between a separable potential like  $V({\bf r})= V\cos(2\pi x/d)\cos(2\pi y/d)$, which gives us a simple square lattice, or more sophisticated non-separable potentials, which give us bipartite lattices (i.e., a lattice with two sublattices) like, for example, the honeycomb lattice.

The displayed Hamiltonian (\ref{2d}) is of fundamental importance in solid-state physics because of its relevance to the integer quantum Hall effect.  However, the exhaustive analysis of the system (\ref{2d}) is still an open problem. For this reason, one usually restricts oneself to the deep quantum regime where the Hamiltonian (\ref{2d}) can be approximated by the tight-binding Hamiltonian. In the case of the simple square lattice and the Landau gauge, this tight-binding Hamiltonian has the form
\begin{align} \label{2d_tb_1}
\widehat{H}=\widehat{H}_0+ d\sum_{\ell,m} \left(F_x \ell + F_y m\right)|\ell,m\rangle\langle \ell,m |  \;, \\
 \label{2d_tb_2}
 \widehat{H}_0=E_0\sum_{\ell,m} |\ell,m\rangle\langle \ell,m |
-\frac{J}{2}\sum_{\ell,m} \left(|\ell+1,m\rangle\langle \ell,m | + h.c.\right) 
-\frac{J}{2}\sum_{\ell,m} \left(e^{i2\pi\alpha\ell}|\ell,m+1\rangle\langle \ell,m | + h.c.\right)  \;,
\end{align}
where $|\ell,m\rangle$ stands for the two-dimensional Wannier states and the parameter $\alpha$, known as the Peierls phase, is given by the ratio of the magnetic flux $\Phi$ through the elementary cell, $\Phi=d^2 B$, to the flux quantum $\Phi_0= hc/e$.
\footnote{Derivation of the tight-binding Hamiltonian (\ref{2d_tb_2}) can be found, for example, in the book \cite{Stoc99}.}
We also mention that available at the Earth conditions magnetic and electric fields correspond to $\alpha\ll 1$ and $dF/J\ll 1$. This justifies further approximations where the energy spectrum of the Hamiltonian $\widehat{H}_0$ is approximated by the multi-degenerate Landau levels, $E_n=\hbar \omega_c(n+1/2)$ with $\omega_c$ being the cyclotron frequency, and the effect of the electric field is taken into account perturbatively by using the linear response theory. 
\begin{figure}[t]
\centering
\includegraphics[width=.6\textwidth]{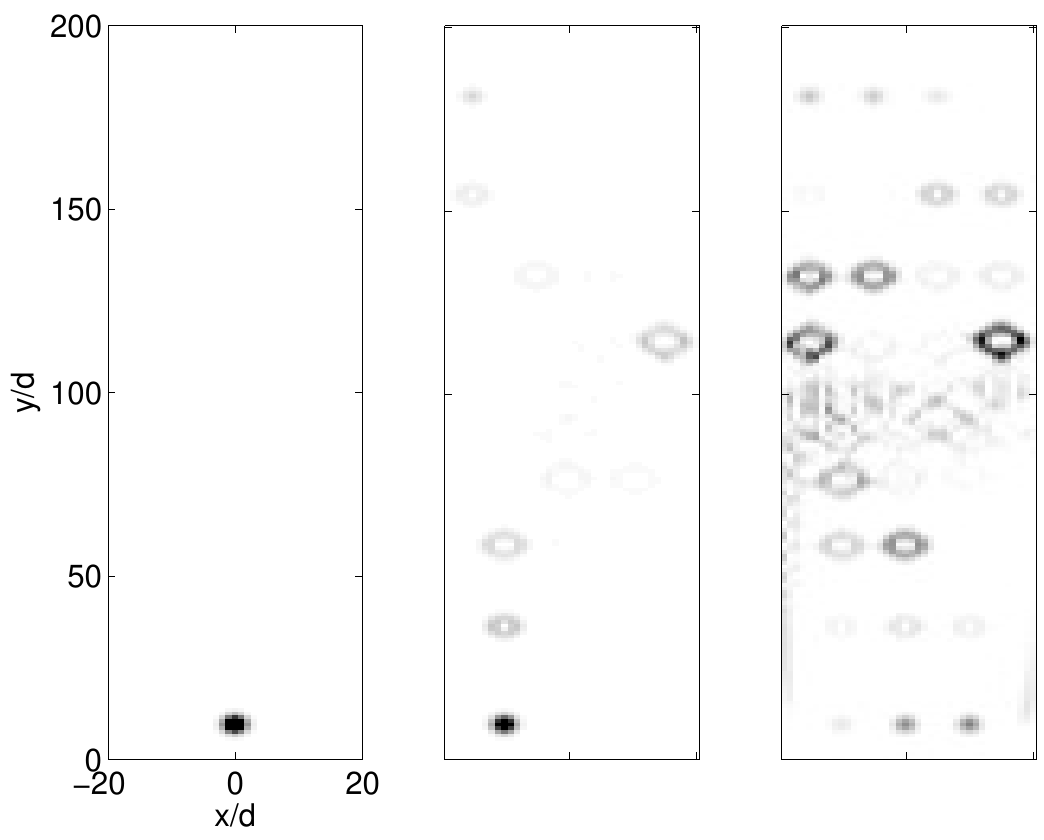}
\caption{Time evolution of a localized wave packet in the square lattice of the width $L=40$ sites induced by an electric field which is alined with the $y$ axis. Snapshots at $t=0$ (left), $t=0.75T$ (middle), and $t=1.5T$ (right) where $T=L/v$ with $v$ being the drift velocity. The system parameters are $J=1$, $\alpha=1/10$, and ${\bf F}=(0,0.02)$. The figure is borrowed from Ref.~\cite{98}.}
\label{fig2_3}
\end{figure}

Recently, interest in Hall physics was renewed due to the appearance of new experimental platforms like cold atoms in a 2D optical lattice, which are subject to the artificial electric and magnetic fields \cite{84,Aide15}. As compared to the crystalline electron, the crucial features are: (i) there are neither impurities nor phonon bath that ensures coherent evolution of the system; (ii) atoms can be made non-interacting that justifies single-particle approach;  and (iii) the cold-atom realization of the tight-binding Hamiltonians (\ref{2d_tb_1}) and (\ref{2d_tb_2}) imposes no restriction on magnitudes of the `electric' and `magnetic' fields.  All these has fundamentally changed the theoretical approach to Hall physics. Indeed, comparing Eq.~(\ref{2d_tb_1}) and  Eq.~(\ref{2d_tb_2}) with  Eq.~(\ref{ws}) and  Eq.~(\ref{tb}) in Sec.~\ref{sec2_2}, it is seen that the discussed Hamiltonian is just a two-dimensional generalization of the Wannier-Stark Hamiltonian, complicated by the presence the Peierls phase $\alpha$. The latter complication, however,  appears to be not crucial and one can develop a theory of the two-dimensional Wannier-Stark states which, if $\alpha\ne0$, we prefer to call Landau-Stark states. In other words, one can relatively easily find both the energy spectrum and eigenfunctions of the Hamiltonian (\ref{2d_tb_1}) \cite{92,85}.

With respect to the chapter subject, our interest in the system (\ref{2d_tb_1}) is because for finite-size lattices or lattices with harmonic confinement, a non-zero $\alpha$ gives rise to chaotic dynamics \cite{98,97}. As an example, Fig.~\ref{fig2_3} shows the wave-packet dynamics of the initially localized packet in the square lattice of the width 40 sites for the `electric' field ${\bf F}$ oriented along the $y$ axis. This dynamics consists of irregular generation of copies of the initial packet at the lattice edges, which then move with the drift velocity $v=F/2\pi\alpha$ perpendicular to the vector ${\bf F}$. Importantly, this dynamics is well described in terms of the Landau-Stark states. It was also shown in Ref.~\cite{98} that the energy spectrum of these states obeys the universal random-matrix (Wigner-Dyson) statistics.

\subsection{Experimental platforms}\label{sec2_7}
%
We described a number of coherent effects, a large part of which were already observed in laboratory experiments with cold atoms in optical lattices. In this respect, i.e., in respect to laboratory studies of  single-particle transport one  should also mention the other popular physical platform -- photonic crystals \cite{photonic}.  Indeed, using photonic crystals one can realize all the  above considered tight-binding Hamiltonians, including the systems (\ref{2d_tb_2}) and (\ref{2d_tb_1}).  And, of course, one has to mention semiconductor superlattices, which 30 years ago were considered to be the main candidate for studying single-particle Quantum Chaos. Each of the listed physical systems has its own advantages and disadvantages. For example, semiconductor superlattices have a rather short coherence time where direct observation of such a simple transport phenomenon like Bloch oscillations  becomes a challenge \cite{Lyss97}. Yet, they are well suited for studying quantum transport (not necessarily coherent) in the presence of a magnetic field. Photonic crystals offer impressive visualization of quantum transport, including  topologically protected transport along the lattice edges \cite{Hafe13}, but they admit a rather short evolution time due to the finite length of the crystal.  Experiments with cold atoms in optical lattices are rather expensive but this is compensated by a number of unique features of the system. In particular, a nice feature of cold-atom systems is a low decoherence rate and practically infinite evolution time, where one can observe thousands of Bloch periods \cite{Gust08}. Beside this, the geometry of optical lattices can be changed continuously which, in particular, allows us to transform a simple lattice into a bipartite lattice and vice versa \cite{Tarr12}. However, the main advantage of cold-atom systems as compared to semiconductor superlattices or photonic crystals is the controllable inter-particle interaction
\footnote{In laboratory experiments the sign (repulsive or attractive) and strength of the atom-atom interaction is changed by applying an external magnetic field. Notice that, since atoms are electrically neutral, this field affects only their internal structure but not their dynamics in a lattice.} 
that opens the door  to many-body physics.

\section{Many-body chaotic dynamics}\label{sec3}
In this section we discuss one of the paradigm models of many-body quantum chaos -- the Bose-Hubbard model. The section is largely based on the review articles \cite{63,103,107} where the reader can find more details together with references to the original research papers. Here we restrict ourselves to the results which we shall need later on when considering dynamics of this model in Sec.~\ref{sec3_5} and the topic of two-terminal transport in Sec.~\ref{sec5}. It should be stressed that here we are not interested in the quantum phase transitions for the ground state, which is a vast but completely different topic in the context of the Bose-Hubbard model \cite{Chan25}.

Formally, the  Bose-Hubbard model is the many-body generalization of the tight-binding Hamiltonian (\ref{tb}) for Bose particles (bosons). In terms of the bosonic creation $\hat{a}_\ell^\dagger$ and annihilation $\hat{a}_\ell$ operators, which creates/annihilates a particle in the Wannier state with the index $\ell$, the Hamiltonian of the Bose-Hubbard model has the form,
\begin{align} \label{BH}
\widehat{\cal H}=E_0 \sum_{\ell=1}^L  \hat{n}_\ell  
  -\frac{J}{2} \sum_{\ell=1}^L \left( \hat{a}^\dag_{\ell+1}\hat{a}_\ell +h.c.\right) +\frac{U}{2}\sum_{\ell=1}^L \hat{n}_\ell(\hat{n}_\ell-1) \;, \quad 
 \hat{n}_\ell=\hat{a}^\dagger_\ell\hat{a}_\ell \;,  \quad [\hat{a}_\ell,\hat{a}_m^\dagger]=\delta_{\ell,m} \;,
 \end{align}
where the last term takes into account collision-like particle-particle interaction. This short-range interaction is typical for neutral cold atoms and for this reason the Bose-Hubbard model  describes very well the cold Bose atoms (for example, $Rb^{87}$) in optical lattices.  Notice, that besides the tunneling rate $J$ and the microscopic interaction constant $U$ the third parameter of the model is the total number of particles  $N$,  which is a conserved quantity. Alternatively, one can use as the third parameter the mean particle density $\bar{n}=N/L$, which will be our choice. For finite $N$ the Hamiltonian (\ref{BH}) is defined in a Hilbert space of dimension 
\begin{displaymath}
{\cal N}=\frac{(N+L-1)!}{N!(L-1)!} 
\end{displaymath}
which is spanned by the Fock states $|{\bf n}\rangle=|n_1,\ldots,n_L\rangle$ where $\sum_{\ell=1}^L n_\ell=N$.
\footnote{In the coordinate representation the Fock states are given by the symmetrized product of the Wannier states. For example, for the two-site Bose-Hubbard model and $N=2$ the state $|1,1\rangle=\left[\Phi_1(x_1)\Phi_2(x_2)+\Phi_1(x_2)\Phi_2(x_1)\right]/\sqrt{2}$.} 

In what follows, if not stated otherwise, we shall consider the Bose-Hubbard chain on a finite interval with periodic boundary conditions.  Notice that with periodic boundary conditions the Bose-Hubbard Hamiltonian  can be rewritten in terms of the operators $\hat{b}_k^\dagger$ and $\hat{b}_k$ which creates/annihilates a particle with quasimomentum of integer index $k$. Then the system Hamiltonian takes the form,
\begin{align} \label{BH_quasimomentum}
\widehat{\cal H}= E_0N - J\sum_{k=1}^L \cos\left(\frac{2\pi k}{L}\right)\hat{b}^\dagger_k\hat{b}_k  +  \frac{U}{2L}\sum_{k_1,k_2,k_3,k_4=1}^L 
     \hat{b}^\dagger_{k_1}\hat{b}^\dagger_{k_2}\hat{b}_{k_3}\hat{b}_{k_4}\delta(k_1+k_2-k_3-k_4)  \;, \quad
     \hat{b}_k=\frac{1}{\sqrt{L}}\sum_{\ell=1}^L \exp\left(i\frac{2\pi k\ell}{L}\right)\hat{a}_\ell \;,
\end{align}
where $\delta$-function (more precisely, a Kronecker symbol) insures conservation of the total quasimomentum.
\begin{figure}
\centering
\includegraphics[width=.6\textwidth]{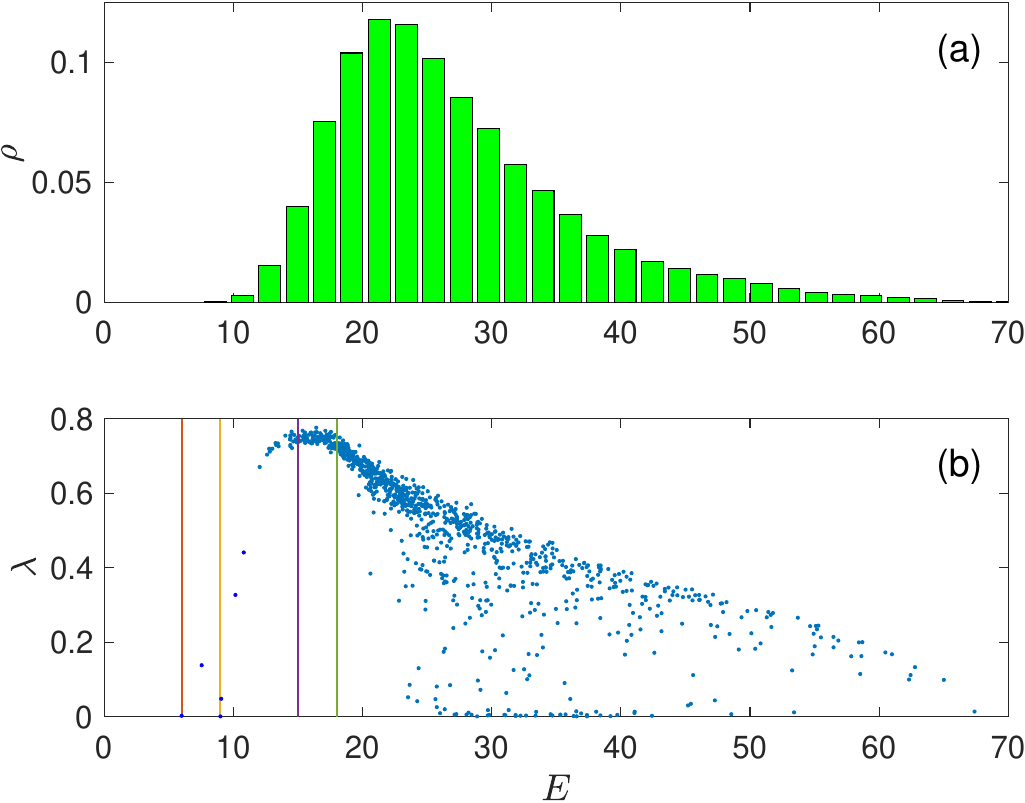}
\caption{Upper panel: Relative volume of the energy shell as a function of the shell energy $E$. Lower panel: Lyapunov exponent $\lambda$ of 1000 trajectories with the initial conditions uniformly distributed over the whole phase space. Vertical lines mark the energies of the periodic trajectories (\ref{nonlinear}). Parameters are $L = 6$, $J = 1$, and $g = 4$. The figure is borrowed from Ref.~\cite{113}.}
\label{fig3_1}
\end{figure}

\subsection{Classical  Bose-Hubbard model}\label{sec3_1}
A remarkable feature of the Bose-Hubbard model is that, unlike the Fermi-Hubbard models, it has the well-defined classical counterpart, 
\begin{align} \label{BH_cl}
H=\Omega \sum_{\ell=1}^L  a^*_\ell a_\ell -\frac{J}{2}\sum_{\ell=1}^L (a^*_{\ell+1}a_\ell + c.c.) + \frac{g}{2}\sum_{\ell=1}^L |a_\ell |^4  \;,  \quad 
\Omega=\frac{E_0}{\hbar} \;,
\end{align}
where $g=U\bar{n}$ is the so-called macroscopic interaction constant. The relation between creation and annihilation operators and classical canonical variables is as follows. First, we use the scaling $\hat{a}_\ell/ \sqrt{\bar{n}}=\tilde{a}_\ell$,  $\hat{a}_\ell^\dagger/ \sqrt{\bar{n}}=\tilde{a}_\ell^\dagger$, $\widehat{\cal H}/\bar{n}=\widetilde{\cal H}$  and then identify the new operators $\tilde{a}_\ell$ and $\tilde{a}_\ell^\dagger$ with the canonical variables $a_\ell$ and $a_\ell^*$. Since the commutator
\begin{align} \label{commute_1}
[\tilde{a}_\ell,\tilde{a}^\dagger_{m}]=\frac{1}{\bar{n}}\delta_{\ell,m}  \;, 
\end{align}
the parameter $1/\bar{n}$ plays the role of an effective Planck constant. Going ahead, we notice that the Hamiltonian (\ref{BH_cl}) can also be  introduced in a rigorous way, see Sec.~\ref{sec3_4} below. 

Let us discuss the classical Bose-Hubbard model in more detail. From the viewpoint of classical mechanics the system (\ref{BH_cl})  is nothing else as a chain of coupled non-linear oscillators. Thus, drawing the analogy  with the famous Fermi-Pasta-Ulam model \cite{Berm05}, one expects that generally it has chaotic dynamics. 
\footnote{The important exception is the two-site Bose-Hubbard system which is integrable. It serves as the model for the Josephson oscillation and the related phenomenon of self-trapping.}
This dynamics takes place in a multi-dimensional phase space where the relative volumes of energy shells determine the density of state of the quantum Bose-Hubbard model \cite{107}. We consider an ensemble of initial conditions randomly distributed over a given energy shell and numerically calculate the trajectories for each initial condition together with their Lyapunov exponents. The Lyapunov exponents are shown in Fig.~\ref{fig3_1} for $L=6$ where one can see that the overwhelming majority of trajectories have positive Lyapunov exponents, i.e., are chaotic. Of course, there are some exceptions, in particular, the so-called nonlinear Bloch waves,
\begin{align} \label{nonlinear}
a_\ell(t)=\exp[i\kappa \ell+iJ\cos(\kappa) t - igt ] \;,\quad \kappa=2\pi k/L \;,
\end{align}
which are formal solutions of the Hamiltonian equation of motion for the system (\ref{BH_cl}).  These solutions are stable if $|\kappa|<\pi/2$  or unstable  if $|\kappa|>\pi/2$.  In the quantum case they correspond to the stable and unstable Bose-Einstein condensates of Bose particles. We mention, in passing,  that in the stable case the trajectory (\ref{nonlinear}) is surrounded by a stability island. Quantizing the latter one obtains the Bogoliubov excitations of the condensate, see Sec.~3.3 in Ref.~\cite{103}. 

\subsection{Quantum Chaos in the  Bose-Hubbard model}\label{sec3_2}
\begin{figure}
\centering
\includegraphics[width=.6\textwidth]{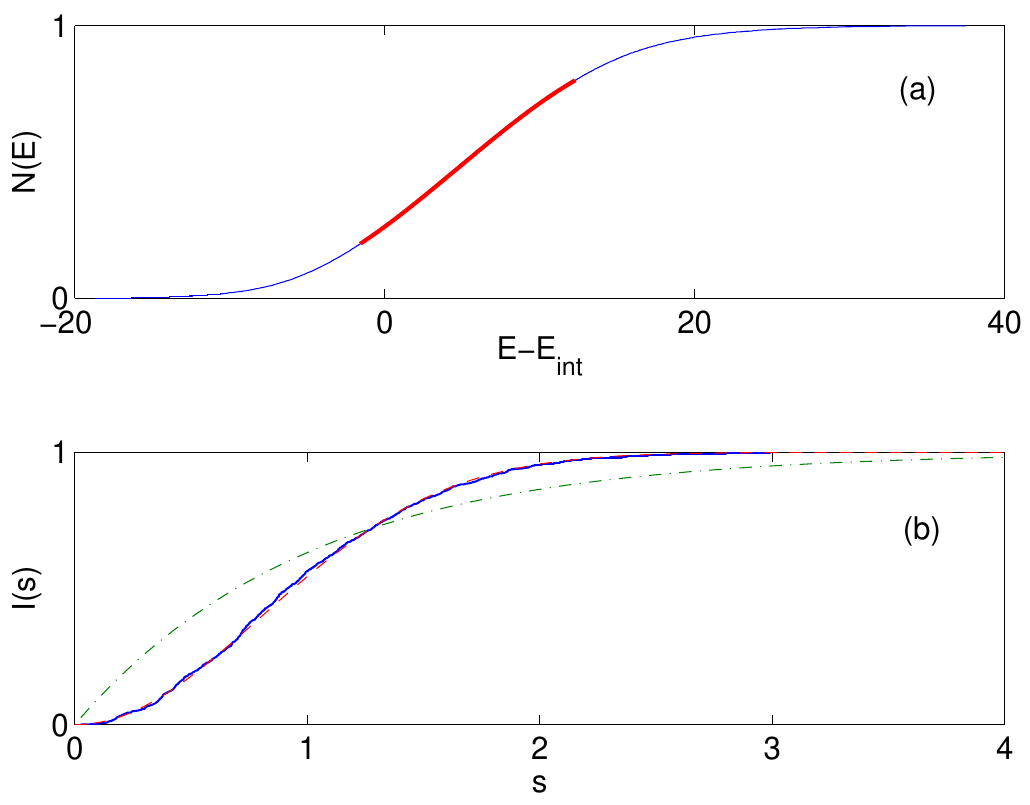}
\caption{Integrated density of states of the 5-site BH model for $g=1$ and $N=19$ and the integrated level-spacing distribution $I(s)$ (solid line in the lower panel) as compared to the integrated Wigner-Dyson distribution (dashed red line) and the integrated Poisson distribution (green dash-dotted line). The energies are taken from the interval marked by the thick red line in the upper panel that comprises 60 percent  from ${\cal N}=1771$ energy levels. The figure is borrowed from Ref.~\cite{107}.} 
\label{fig3_3}
\end{figure}

First of all we note that the Bose-Hubbard model is integrable in the case $J=0$, where the system eigenstates are the (position) Fork number states, and in the case $U=0$, where the system eigenstates are the quasimomentum Fock states given by the symmetrized product of the single-particle Bloch states. In the general case, based on the previous classical analysis,  the Bose-Hubbard system is expected to be chaotic in the sense of quantum chaos. In other words, its energy spectrum and eigenfunctions should possess universal properties similar to those of random matrices of the corresponding symmetry class. For details of the statistical analysis of the energy spectrum of the Hamiltonian (\ref{BH}) we refer the reader to Refs.~\cite{66,Koll10,Sant10,Paus21}. Here we only mention that the transition to chaos, which is manifested in the transition from Poisson to Wigner-Dyson statistics (see Fig.~\ref{fig3_3}), is a crossover over the interaction strength $U$. In the case $\bar{n}\sim 1$   this requires $U\sim J$ 
\footnote{The transition to chaos for $\bar{n}=1$ is studied in full detail in  Ref.~\cite{Paus21}.}
and  in the semiclassical case $\bar{n}\gg1$ the region of chaos in the system parameter space spanned by the energy $E$ and the ratio $g/J$ can be well identified by using the Lyapunov analysis of the classical system (\ref{BH_cl}).

Along with the statistical analysis of the energy spectrum  the respective analysis of the system eigenstates is no less important \cite{Sant10,Beug14,Beug15,Schl16}.  In particular, the work \cite{Schl16} analyses eigenstates $| \Psi_E \rangle$ of the Bose-Hubbard model from the viewpoint of Dynamical Thermalization Conjecture {\cite{Borg16,chapter_18}}. Roughly, this conjecture states that the occupation numbers of the quasimomentum states, $\langle n_k \rangle=\langle \Psi_E | \hat{b}^\dagger_k\hat{b}_k | \Psi_E \rangle$ obey the Bose-Einstein distribution,
\begin{align} \label{einstein}
\langle n_k \rangle=\frac{1}{e^{\beta(E_k-\mu)}-1} \;, \quad E_k=-J\cos\left(\frac{2\pi k}{L}\right) + U\bar{n} \;,
\end{align}
where the inverse temperature $\beta$ and the chemical potential $\mu$ are uniquely determined by the eigenstate energy $E$ and the mean particle density $\bar{n}$. A number of numerical results in favor of this conjecture were reported. These results suggest that the Bose-Hubbard system can serve as a thermodynamic bath. In its turn, the latter hypothesis was tested (i) in Ref.~\cite{61}, which studies Bloch oscillations of Fermi atoms colliding with Bose atoms; (ii) in Ref.~\cite{117}, which analyzes the dynamics of a two-level system weakly coupled to a finite-size quantum Bose-Hubbard model; and (iii) in the recent work Ref.~\cite{preprint2}, which discusses self-thermalization dynamics of the model by using a pseudoclassical approach.  

\subsection{Single-particle density matrix}\label{sec3_3}
In what follows we shall often refer to the single-particle density matrix as SPDM.  This matrix has size $L\times L$ with the matrix elements 
\begin{align} \label{spdm}
\rho_{\ell,m}(t)=\langle\Psi(t)| \hat{a}_\ell^\dagger \hat{a}_m |\Psi(t)\rangle \;.
\end{align}
Although the SPDM does not contain the whole information which is encoded in the many-body wave-function, the knowledge of  $\hat{\rho}(t)$ suffices to calculate single-particle observables like, for example, the mean current/velocity of an ensemble of Bose particles,
\begin{align} \label{mean_current_many}
j(t)=\langle \Psi(t)|\widehat{\cal J} |\Psi(t)\rangle \;, \quad  
\widehat{\cal J}=\frac{j_0}{2i} \sum_{l=1}^{L} ( \hat{a}^\dag_{\ell+1}\hat{a}_\ell - h.c. ) \;.
\end{align}
Using the SPDM Eq.~(\ref{mean_current_many}) simplifies to
\begin{align} \label{mean_current_spdm}
j(t)={\rm Tr}[\hat{j}\hat{\rho}(t)] \;, \quad  
j_{\ell,m}=\frac{j_0}{2i} (\delta_{\ell,m+1}-\delta_{\ell+1,m}) \;.
\end{align}
The matrix $\hat{\rho}$ (more exactly, its entropy $S=-{\rm Tr}[\hat{\rho}\log \hat{\rho}]$)  can also serve as indicator of entanglement  of the many-body state $|\Psi_E\rangle$. However, the most useful property of the SPDM is that it can be approximately calculated without knowing  the full many-body wave function  $|\Psi(t)\rangle$. One of these approximate methods is the pseudoclassical approach which we discuss in the next subsection. 

\subsection{Pseudoclassical approach}\label{sec3_4}
The pseudoclassical approach uses the phase-space presentation for quantum operators. Given $|\Psi(t)\rangle$ to be the many-body wave function we define the Husimi function 
\begin{equation}
\label{husimi}
f_{qu}({\bf a},{\bf a}^*,t)=|\langle {\bf a} |\Psi(t)\rangle |^2 \;, \quad
 |{\bf a}\rangle= \frac{1}{\sqrt{N!}}\left(\sum_{\ell=1}^L a_\ell \hat{a}^\dagger_\ell \right)^N |vac\rangle \;, \quad
 \sum_{\ell=1}^L | a_\ell |^2=1 \;,
\end{equation}
where $|{\bf a}\rangle$ are the so-called Schwinger coherent state for $SU(L)$. Notice that the Husimi function (\ref{husimi}) is a function of $L$ complex amplitude $a_\ell$ or $a_\ell^*$ and time. In terms of the Husimi function (\ref{husimi}) the Schr\"odinger equation for the wave function $|\Psi(t)\rangle$ takes the form 
\begin{equation}
\label{liuvil_qu}
\frac{\partial f_{qu}}{\partial t}=    \{ H,f_{qu} \} + O\left(\frac{1}{\bar{n}}\right) \;, \quad
 \{H,f\} \equiv  -i\sum_{\ell=1}^L \left( \frac{\partial H}{\partial a_\ell}\frac{\partial f}{\partial a_\ell^*}
  - \frac{\partial H}{\partial a_\ell^*}\frac{\partial f}{\partial a_\ell} \right) \;,
\end{equation}
where the $c$-number Hamiltonian $H$ is given in Eq.~(\ref{BH_cl}), and we refer the reader to the work \cite{Trim08} for the explicit form of terms which are inverse proportional to $\bar{n}$.  As mentioned in Sec.~\ref{sec3_1}, the Hamiltonian (\ref{BH_cl}) formally follows from the microscopic Hamiltonian (\ref{BH}) by using `quantization rules'  $a_\ell \leftrightarrow \hat{a}_\ell/\sqrt{\bar{n}}$, $a^*_\ell\leftrightarrow \hat{a}^\dagger_\ell/\sqrt{\bar{n}}$, and $H \leftrightarrow \widehat{\cal H}/\bar{n}$. The approach using the Husimi function provides justification for these heuristic quantization rules. 

The essence of the pseudoclassical method is that for $\bar{n}\gg 1$ we can approximate the Husimi function by the classical distribution function obeying the Liouville equation,  
\begin{equation}
\label{liuvil_cl}
\frac{\partial f_{cl}}{\partial t}=\{H,f_{cl}\}  \;.
\end{equation}
Then elements of the single-particle density matrix are found by calculating the integral from the function $a_\ell^*a_m f_{cl}({\bf a},{\bf a}^*;t)$ over the whole phase-space. Furthermore, in the case where we cannot solve Eq.~(\ref{liuvil_cl}) analytically, we can always resort to the Monte-Carlo simulation where elements of the single-particle density matrix are found by solving the Hamiltonian equation of motion,
\begin{align} \label{eq_H}
i\dot{a}_\ell=\frac{\partial H}{\partial a^*_\ell} = \Omega a_\ell-\frac{J}{2}\left(a_{\ell+1} +a_{\ell-1}\right) + g|a_\ell |^2 a_\ell \;,
\end{align}
and then averaging the product $a_\ell^*(t)a_m(t)$ over the ensemble of initial conditions,
\begin{align} \label{average}
\rho_{\ell,m}(t)=\langle a_\ell^*(t) a_m(t) \rangle \;.
\end{align}
The corresponding ensemble of initial conditions is obviously determined  by the Husimi function (\ref{husimi}) at $t=0$ which, in its turn, is determined by the initial wave-function. For this reason it is named the {\em quantum ensemble}. In the numerical implementation  of the pseudoclassical approach the generating of the quantum ensemble of classical initial conditions is the most difficult procedure \cite{103}.

We conclude this subsection with two remarks. First, the above discussed quantization rule, where $1/\bar{n}$ plays the role of an effective Planck constant,  can be used to find the spectrum of the Bose-Hubbard model from its classical counterpart. For the two-site Bose-Hubbard model, which is an integrable system, this was demonstrated in Ref.~\cite{Grae07}.  It was shown that  the semiclassical quantization perfectly reproduces the system energy spectrum $E_n=E_n(U)$ already for $\bar{n}=5$. In principle, by using the Gutzwiller periodic orbit theory,  one can try to find the spectrum of the $L$-site Bose-Hubbard model. The semiclassical analysis (not to be confused with pseudoclassical analysis) of the $L$-site Bose-Hubbard model is discussed in chapter \cite{chapter_18} in this volume.

The second remark concerns the open Bose-Hubbard model which will be considered in Sec.~\ref{sec5}. The open Bose-Hubbard model does not conserve the number of particles $N$ and, hence, it is defined in the extended Hilbert space given by the sum of the finite-size Hilbert spaces associated with a given  $N$. In this case one defines the Husimi function by using the Glauber coherent states instead of Schwinger $SU(L)$ states. Alternatively, we can use  the Wigner function instead of the Husimi function,
\begin{align} \label{wigner}
f_{qu}(a,a^*;t)=\frac{1}{\pi^2}\int {\rm d}\beta{\rm d}\beta^*e^{-i(a\beta+a^*\beta^*)} 
{\rm Tr}\left[e^{i(\beta\hat{a}+\beta^*\hat{a}^\dagger)} {\cal R}(t) \right] \;,\quad
{\cal R}(t)=|\Psi(t)\rangle\langle \Psi(t) | \;,
\end{align}
which is displayed here for the one-site Bose-Hubbard model. For both functions the classical distribution function obeying the Liouville equation (\ref{liuvil_cl}) serves as the zero order approximation,  although there is an unwanted freedom in defining the effective Planck constant.  In spite of this problem, the pseudoclassical approach proved to be a powerful tool for analyzing  the quantum dynamics of the open Bose-Hubbard model as well.
\begin{figure}
\centering
\includegraphics[width=.45\textwidth]{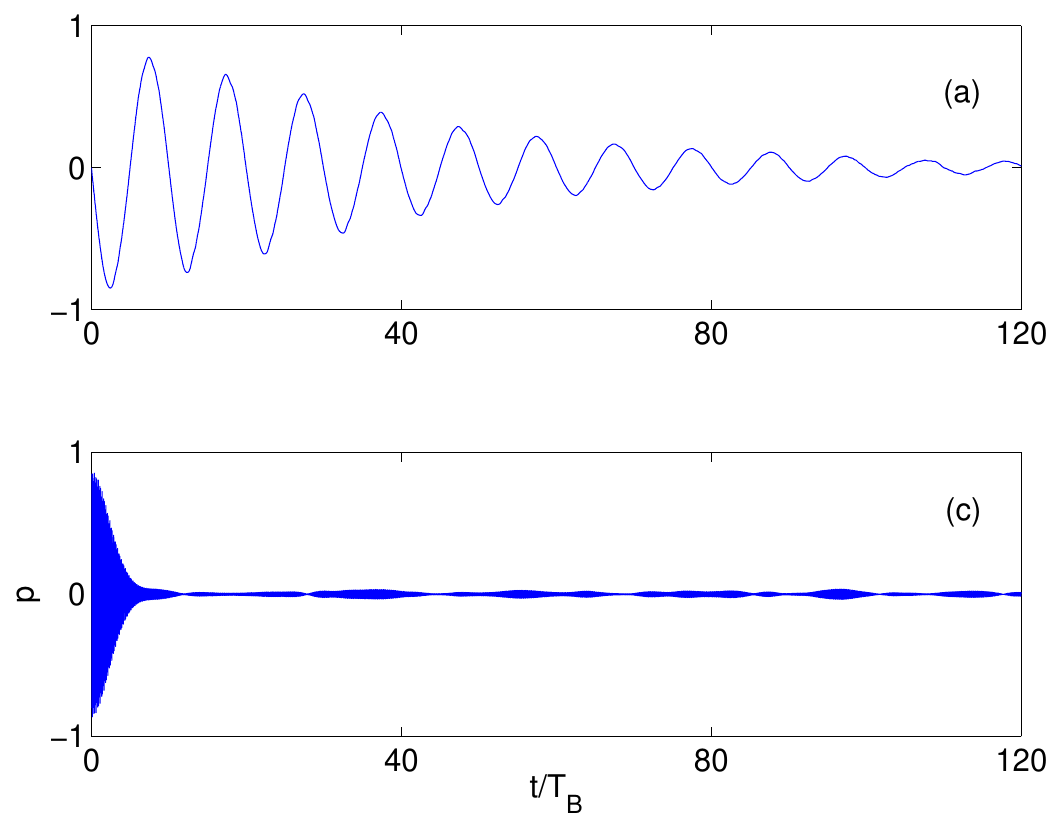}
\includegraphics[width=.45\textwidth]{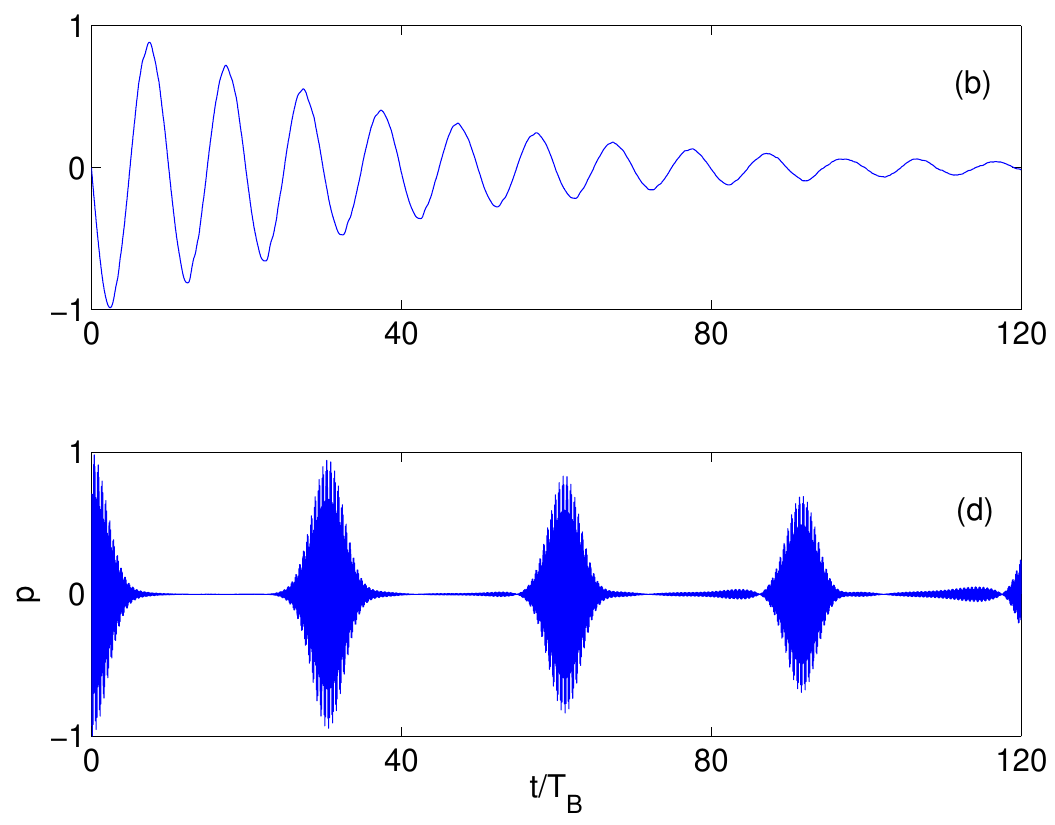}
\caption{Bloch oscillations of interacting atoms calculated by using the pseudoclassical approach, panels (a) and (c), and the fully quantum-mechanical simulations, panels (b) and (d). The system parameters are $N = 15$, $L = 5$, $J = 1$, $U = 0.1/3$, and $F = 0.1$, panels (a) and (b), and $F = 10$, panels (c) and (d). The figure is borrowed from Ref.~\cite{103}.}
\label{fig3_2}
\end{figure}

\subsection{Bloch oscillation of interacting atoms}\label{sec3_5}
As the first example of practical applications of the pseudoclassical approach we consider the simplest many-body transport problem, namely, Bloch oscillations of $N$ interacting atoms in an optical lattice.  As the initial condition we consider the maximally coherent state
\begin{align} \label{bec}
|\Psi(t=0)\rangle=\frac{1}{\sqrt{N!}} \left(\hat{b}^\dagger_{k=0}\right)^N | vac \rangle \;, \quad 
\hat{b}^\dagger_{k=0}=L^{-1/2}\sum_\ell \hat{a}_\ell^\dagger \;.
\end{align}
which is the Bose-Einstein condensate of cold atoms in the zero quasimomentum state. We shall analyze Bloch oscillations in momentum space which allows us to use the periodic boundary condition.  Indeed, by using a simple substitution $a_\ell(t)=a'_\ell(t)\exp(-iF\ell t)$ the static force $F$ is moved into the kinetic energy term as an oscillating phase. This  preserves the translational symmetry of the system but makes the system Hamiltonian explicitly time-dependent,    
\begin{equation}
\label{gauge}
H(t)=\Omega\sum_{\ell=1}^L a^*_\ell a_\ell-\frac{J}{2}\sum_{\ell=1}^L \left(a^*_{\ell+1}a_\ell e^{iFt} + c.c.\right) 
+ \frac{g}{2}\sum_{\ell=1}^L |a_\ell |^4  \;, \quad g=U\bar{n} \;.
\end{equation}
(here we omit prime for the new canonical variables).

According to the Bloch acceleration theorem, the static force $F$ changes the quasimomentum linearly in time. After one quarter of the Bloch period this brings the system into a parameter region where the nonlinear Bloch wave Eq.~(\ref{nonlinear}) is unstable and, thus, all neighboring trajectories are chaotic.  Because of the exponential sensitivity of chaotic trajectories to initial conditions this leads to exponential decay of the off-diagonal elements of the SPDM (\ref{average}). As a consequence, the momentum-space Bloch oscillations irreversibly decay in the course of time, see Fig.~\ref{fig3_2}(a).  Remarkably, the pseudoclassical approach perfectly reproduces the result of the exact numerical simulation of the system dynamics, compare panels (a) and (b) in Fig.~\ref{fig3_2}. It should be mentioned, however,  that this is not always the case. In particular, one can stabilize dynamics of the system (\ref{gauge}) by increasing $F$ \cite{103}. Then, in the quantum case the irreversible decay of Bloch oscillation changes to decay with revivals, see Fig.~\ref{fig3_2}(d). Since revivals are an interference effect, the pseudoclassical approach is incapable to reproduce it, see Fig.~\ref{fig3_2}(c). In this sense the chaotic dynamics of the system is crucial for both the exponential decay of Bloch oscillations and validity of  the pseudoclassical approach \cite{103}.

Let us add two important remarks. First, both regimes of irreversible decay and decay with revivals were observed in laboratory experiments Ref.~\cite{Mein14} on Bloch oscillations in interacting rubidium atoms. Thus, one may consider the chaotic nature of the Bose-Hubbard model as an experimentally confirmed result.

Second, in the chaotic regime the numerically observed dynamics of the single-particle density matrix  $\hat{\rho}(t)$ coincides  with solution of the following master equation 
\begin{align} \label{master_spdm} 
\frac{d \hat{\rho}}{dt}=-i[\widehat{H}(t),\hat{\rho}] +\widetilde{\cal L}(\hat{\rho}) \;, \quad  \hat{\rho}(t=0)=| \psi_{k=0}\rangle\langle \psi_{k=0} | \;,
\end{align}
where $\widehat{H}(t)= \Omega\sum_\ell |\ell\rangle\langle \ell | - (J/2)\sum_\ell (|\ell+1\rangle\langle \ell | e^{iFt} + h.c.)$ and
\begin{equation}\label{decoherence}
\widetilde{\cal L}(\hat{\rho})_{\ell,m}= -\eta(1-\delta_{\ell,m})\rho_{\ell,m} \;.
\end{equation}
Notice that the operator (\ref{decoherence}) can be rewritten in the Lindblad form as
\begin{equation}
\widetilde{\cal L}(\hat{\rho})= -\frac{\eta}{2} \sum_{\ell=1}^L 
( \hat{\ell}\hat{\ell}\hat{\rho}-2\hat{\ell}\hat{\rho}\hat{\ell}+\hat{\rho}\hat{\ell}\hat{\ell} )  \;,\quad  \hat{\ell}=| \ell \rangle\langle \ell | \;,
\end{equation}
which justifies the term `local dephasing'. Usually the master equation (\ref{master_spdm})  appears in problems where the system of interest is coupled to a bath. Thus,  in the chaotic regime the Bose-Hubbard model is a bath for itself. Based on this result we  put forward a hypothesis that in terms of  SPDM the effect of a weak inter-particle interaction can be mimicked by the local dephasing operator (\ref{decoherence}).

\subsection{Concluding remarks}

Through the section we consider only the Bose-Hubbard model. Of course, there are other models of many-body Quantum Chaos in the physical literature. The most popular with respect to quantum transport are spin chains and interacting spinless fermions in one-dimensional lattices \cite{Bert21}. Compared to interacting bosons, the case of Fermi particles is simpler from the numerical viewpoint because of a smaller dimension of Hilbert space. However,  Fermi systems have no well-defined classical counterparts and, thus, they require genuine quantum methods of analysis. The spin systems place themselves between fermionic and bosonic systems. Indeed, for spin-$1/2$ chains we have a one to one correspondence with spinless fermions in a one-dimensional lattice, while spin chains with a large spin are closer to bosonic systems. We would also like to stress that any numerical proof that any particular many-body system is chaotic in the sense of quantum chaos is mainly of academic interest and it is more important to find physical manifestation of this fact, which can be detected  in a laboratory experiment. As concerns the Bose-Hubbard model, we have at least one clear manifestation -- the irreversible decay of Bloch oscillations of interacting Bose atoms. Also, signatures of chaos can be seen in the quench dynamics of the system where one follows the time evolution of the initial state which is {\em not} the eigenstate of the Bose-Hubbard Hamiltonian. This problem was analyzed in the theoretical work Ref.~\cite{Paus25} which also contains references to laboratory experiments on the quench dynamics of the cold Bose atoms in optical lattices.

\section{Two-terminal transport} \label{sec4}
In this and the remaining sections we study quantum transport in a two-terminal setup. The term  `two-terminal' means that we have two reservoirs of quantum  particles which are connected by a  lattice of the length $L$. We are interested in the particle current across the lattice. Some questions to ask are how the current depends on the strength of inter-particle interaction, on the lattice length $L$, on the intensity of the decoherence process which may be present in the system, and so on. Until quite recently these questions were addressed exclusively with respect to electrons in a solid crystal  but nowadays they are readdressed with respect to cold atoms in optical potentials \cite{Krin17,Corm19} and to microwave photons in superconducting circuits \cite{Fitz17,Fedo21}. In this brief review it is impossible to mention all theoretical approaches  to the two-terminal transport. Because of this we restrict ourselves to a single approach -- the master equation for the reduced density matrix of bosonic particles in a lattice.

\subsection{Markovian master equation for the reduced density matrix} \label{sec4_1}
Let us consider two reservoirs of Bose particles in  thermal equilibrium Eq.~(\ref{einstein}) which we display  here one more time,
\begin{align} \label{dirac}
\langle n_k \rangle=\frac{1}{e^{\beta(E_k-\mu)}-1}  \;.
\end{align}
In Eq.~(\ref{dirac}) $\beta$ is the inverse temperature, $\mu$ the chemical potential (which together with $\beta$ determines the mean particle density $\bar{n}$ in the respective reservoir), $E_k$ are the energies of the single-particle reservoir states, and $\langle n_k \rangle\equiv\bar{n}_k$ the occupation numbers of these states. For the sake of simplicity we assume equal temperatures for both reservoirs. Also we shall assume local coupling between the system and reservoirs, i.e., the particle can tunnel from the left reservoir to the first site of the lattice and from the right reservoir to the last site of the lattice. 
Of course, the reverse process is also possible. This assumption fixes the explicit form of the interaction terms $\epsilon \widehat{\cal H}_{int}^{({\rm i})}$ in the total Hamiltonian of the composed system, 
\begin{align} \label{total}
\widehat{\cal H}_{tot}=  \widehat{\cal H} +  \sum_{\rm i=L,R} \left(\widehat{\cal H}_r^{({\rm i})} 
+ \epsilon \widehat{\cal H}_{int}^{({\rm i})} \right) \;,
\end{align}
where the index ${\rm i}$ labels the left and right reservoirs and we introduce the coupling constant $\epsilon$. Next, it is assumed  that the total density matrix of the system ${\cal R}_{tot}(t)$ obeys the quantum Liouville equation $d {\cal R}_{tot}/dt=-i[\widehat{\cal H}_{tot}, {\cal R}_{tot}]$. Finally, by using the standard procedure which involves the Born-Markov approximation 
\footnote{Loosely speaking, the Born approximation assumes that our system of interest does not affect the thermodynamic properties of a bath. Markov approximation assumes a rapid decay of correlations, i.e., the absence of memory effects. If Markov approximation is not justified, the resulting master equation for the reduced density matrix becomes an integro-differential equation. We come back to this point later on in Sec.~\ref{sec4_4}.}     
one traces out reservoirs degrees of freedom and obtains the master equation for the reduced density matrix ${\cal R}(t)$ of the carriers in the lattice,
\begin{align} \label{markovian_many}
\frac{d {\cal R}}{d t}=-i[\widehat{\cal H},{\cal R}] + \sum_{\rm i=L,R} {\cal L}_{\rm i}({\cal R})   \;.
\end{align}
In Eq.~(\ref{markovian_many}) the Lindblad operators  ${\cal L}_{\rm L}({\cal R} )$ and  ${\cal L}_{\rm R}({\cal R} )$ take into account the process of particle exchange between the lattice and reservoirs and are parametrized by  the mean particle densities $\bar{n}_{\rm i}$ in reservoirs.  Since the coupling was assumed to be local, the explicit form of these Lindblad operators is the following,
\begin{align} \label{lindblad}
{\cal L}_{\rm i}({\cal R})=-\frac{\tilde{\gamma}}{2} \left[(\bar{n}_{\rm i}+1)
\left(\hat{a}_{\rm i}^{\dagger}\hat{a}_{\rm i} {\cal R }-2\hat{a}_{\rm i}{\cal R }\hat{a}_{\rm i}^{\dagger}
+{\cal R }\hat{a}_{\rm i}^{\dagger}\hat{a}_{\rm i} \right)  
+ \bar{n}_{\rm i}
\left(\hat{a}_{\rm i}\hat{a}_{\rm i}^{\dagger}{\cal R }-2\hat{a}_{\rm i}^{\dagger}{\cal R }\hat{a}_{\rm i}
+{\cal R }\hat{a}_{\rm i}\hat{a}_{\rm i}^{\dagger} \right) \right] \;,
 \end{align}
where $\tilde{\gamma}\sim\epsilon^2$ is the particle exchange rate. In Eq.~(\ref{lindblad}) we use the convention $\hat{a}_{\rm i= L}\equiv \hat{a}_{1}$ and $\hat{a}_{\rm i=R}\equiv \hat{a}_{L}$ and the same notations are used for the creation operators.  It is easy to show that the operator ${\cal L}_{\rm L}( {\cal R} )$ `tries' to equilibrate population of the first lattice site with the mean particle density $\bar{n}_{\rm L}$ in the left reservoir  while the operator  ${\cal L}_{\rm R}( {\cal R} )$ tries to equilibrate population of the last lattice site with the mean particle density $\bar{n}_{\rm R}$ in the right reservoir. If $\bar{n}_{\rm R}\ne \bar{n}_{\rm L}$ the system is frustrated which leads to a non-equilibrium steady state (NESS) with non-zero particle current between the two reservoirs. 
\footnote{The case of a quasi-stationary NESS, which is the typical situation for a cold-atom laboratory experiment, where the number of atoms is large ($\sim 10^5$) but finite,  is considered in Ref.~\cite{Ama20,preprint2}.}

First, we consider the case of non-interacting bosons where the Hamiltonian $\widehat{\cal H}$ is a quadratic function of creation and annihilation operators, 
\begin{align} \label{tb_many}
\widehat{\cal H}=E_0 \sum_{\ell=1}^L  \hat{a}^\dag_{\ell}\hat{a}_\ell  - \frac{J}{2} \sum_{\ell=1}^L \left( \hat{a}^\dag_{\ell+1}\hat{a}_\ell +h.c.\right) \;. 
\end{align}
In this case the master equation (\ref{markovian_many}) can be rewritten in terms of the SPDM $\rho_{\ell,m}(t)={\rm Tr}[\hat{a}_\ell^\dagger\hat{a}_m {\cal R}(t)]$ which, as already mentioned in Sec.~\ref{sec3_3}, suffices to calculate  the particle current across the lattice.  Multiplying both parts of Eq.~(\ref{markovian_many}) by  $\hat{a}_\ell^\dagger\hat{a}_m$ and taking the trace we obtain after some algebra
\begin{align} \label{markovian_spdm}
\frac{d \hat{\rho}}{d t}=-i[\widehat{H},\hat{\rho}] +\sum_{\rm i=L,R}{\cal L}_{\rm i}(\hat{\rho})  \;,
\end{align}
where $\widehat{H}$ is  the tight-binding Hamiltonian (\ref{tb}) and  the single-particle Lindblad operators have the form  
\begin{align} \label{lindblad_spdm}
{\cal L}_{\rm i}(\hat{\rho})_{\ell,m}=-\frac{\tilde{\gamma}}{2}\rho_{\ell,m} (\delta_{\ell,{\rm i}}+\delta_{{\rm i},m})
+\tilde{\gamma}\bar{n}_{\rm i}\delta_{\ell,{\rm i}}\delta_{m,{\rm i}}  
\end{align}
(here, as before, the index ${\rm i}$ implies the first lattice site if  ${\rm i=L}$ and the last lattice site if ${\rm i=R}$). 
If we are interested only in the stationary regime, the differential equation (\ref{markovian_spdm}) is substituted by the algebraic equation for the stationary matrix $\bar{\rho}$,  
\begin{align} \label{algibraic_spdm}
\sum_{\rm i=L,R}{\cal L}_{\rm i}(\bar{\rho}) - i[\widehat{H},\bar{\rho}] =0 \;.
\end{align}
The latter equation can be solved analytically by using a three-diagonal ansatz, i.e., only the elements on the main diagonal and the nearest neighbor diagonals differ from zero \cite{Ivan13,112}. Diagonal elements $\bar{\rho}_{\ell,\ell}$ give the stationary populations of the lattice sites and off-diagonal elements  $\bar{\rho}_{\ell,\ell\pm 1}$, which appear to be pure imaginary, determine the stationary current $\bar{j}$ across the lattice. The resulting equation for the stationary current has an amazingly simple form,
 \footnote{We mention the similar equations for the spin current in boundary-driven Heisenberg spin chains \cite{Kare09,Bert21,Land22}.}
\begin{align} \label{current}
\bar{j}=\frac{J^2\tilde{\gamma}}{J^2+\tilde{\gamma}^2} \frac{\bar{n}_{\rm L}-\bar{n}_{\rm R}}{2} \;,
\end{align}
which perfectly fits our intuition that the current should be proportional to the difference in the reservoir particle densities. We remind the reader that these densities are uniquely determined by the reservoirs temperature $1/\beta$ and their chemical potentials $\mu_{\rm L}$ and $\mu_{\rm R}$. Considering situation where  $\mu_{\rm L}=\mu+\Delta\mu/2$ and  $\mu_{\rm R}=\mu-\Delta\mu/2$ we define the lattice conductance $\sigma=\sigma(\mu)$ as 
\begin{align} \label{conductance}
 \sigma(\mu)=\lim_{\Delta\mu\rightarrow 0} \frac{\bar{j}(\mu;\Delta\mu)}{\Delta\mu} \;.
 \end{align}

\subsection{The case of fermionic carriers} \label{sec4_6}
In the case of Fermi particles all bosonic creation and annihilation operators in the displayed equations are substituted by the fermionic operators. Then, because of the different commutation relations, the prefactor $(\bar{n}_{\rm i}+1)$ in the Lindblad operator (\ref{lindblad}) is replaced by the prefactor  $(1-\bar{n}_{\rm i})$. This, however, changes neither the master equation (\ref{markovian_spdm}) nor the result (\ref{current}) for the stationary current. The only difference is that for bosonic carriers the mean particle densities $\bar{n}_{\rm L}$ and $\bar{n}_{\rm R}$ are arbitrary non-negative numbers while for fermions they cannot exceed unity. 

Do the master equations (\ref{markovian_many}) or (\ref{markovian_spdm}) describe the whole physics of two-terminal transport of non-interacting Bose or Fermi particles? Expectedly, not. In particular, these master equations do not describe the phenomenon of resonant transmission \cite{Datt95}. The latter is the mechanism behind the observation that  the conductance $\sigma(\mu)$ is generally not a monotonic function of $\mu$, as stated in  Eq.~(\ref{current}),  but shows a number of transmission peaks. The reason for the failure of  describing  the resonant transport within the framework of the master equation (\ref{markovian_many}) is the Markov approximation. This  approximation assumes that the system-reservoir correlations, which are due to quantum entanglement between the reservoirs and the system states \cite{117}, decay during a characteristic time $\tau$ which is the shortest time in the problem. This is not always the case. We discuss the situation where the time $\tau$ is not the shortest time in the next two subsections by using a more elaborated model for the two-terminal transport.  
For pedagogical reasons, namely, to relate the master equation approach to the Landauer theory  \cite{Datt95} we temporally switch to the case of fermionic carriers. It should be remembered, however, that all results are equally applied to bosonic carriers. These results  will give us a reference point when considering the case of interacting bosons in Sec.~\ref{sec5}. Besides this, the following analysis of two-terminal transport clarifies the  meaning of the Born-Markov approximation which, in fact, are two different approximations and it would be more correct to call it the Born and Markov approximations.

\subsection{Semi-microscopic model of particle reservoirs} \label{sec4_2}
The model discussed in the present subsection was first introduced in Ref.~\cite{Ajis12} and then independently reintroduced in Ref.~\cite{Grus16} and Ref.~\cite{120}. We shall use the variant and notations of Ref.~\cite{120}. The key point of the model is that it takes into account the finite thermalization rate  for particles in the reservoirs.  In other words, if we push a reservoir out of thermal equilibrium, it will take a finite time till it comes back to thermal equilibrium. In the semi-microscopic approach this process of thermalization is mimicked  by using the following master equation for the reservoir density matrix ${\cal R}_r$,
\begin{align} \label{master_bath}
\frac{d  {\cal R}_r}{d t}=-i[ \widehat{\cal H}_r, {\cal R}_r] + {\cal L}( {\cal R}_r) \;, 
\quad   \widehat{\cal H}_r =\sum_k E_k \hat{b}_k^{\dagger}\hat{b}_k  \;, \\
{\cal L}({\cal R}_r)=-\frac{\gamma}{2} \sum_k \left[(1-\bar{n}_k)
\left(\hat{b}_k^{\dagger}\hat{b}_k {\cal R}_r-2\hat{b}_k{\cal R}_r\hat{b}_k^{\dagger} +{\cal R}_r\hat{b}_k^{\dagger}\hat{b}_k \right)  
+ \bar{n}_k
\left(\hat{b}_k\hat{b}_k^{\dagger}  {\cal R}_r-2\hat{b}_k^{\dagger} {\cal R}_r\hat{b}_k 
+ {\cal R}_r\hat{b}_k\hat{b}_k^{\dagger} \right) \right] \;, 
 \end{align}
where the index $k$ labels the reservoir eigenstates and occupation numbers $\bar{n}_k$ are given by the Fermi-Dirac distribution. (The Fermi-Dirac distribution differs from the Bose-Einstein distribution (\ref{dirac}) by the plus sign in the denominator instead of the minus sign.)  The master equation  (\ref{master_bath}) ensures relaxation of the reservoir to the thermal state within the characteristic time $\tau\sim1/\gamma$. 
\begin{figure}
\centering
\includegraphics[width=.4\textwidth]{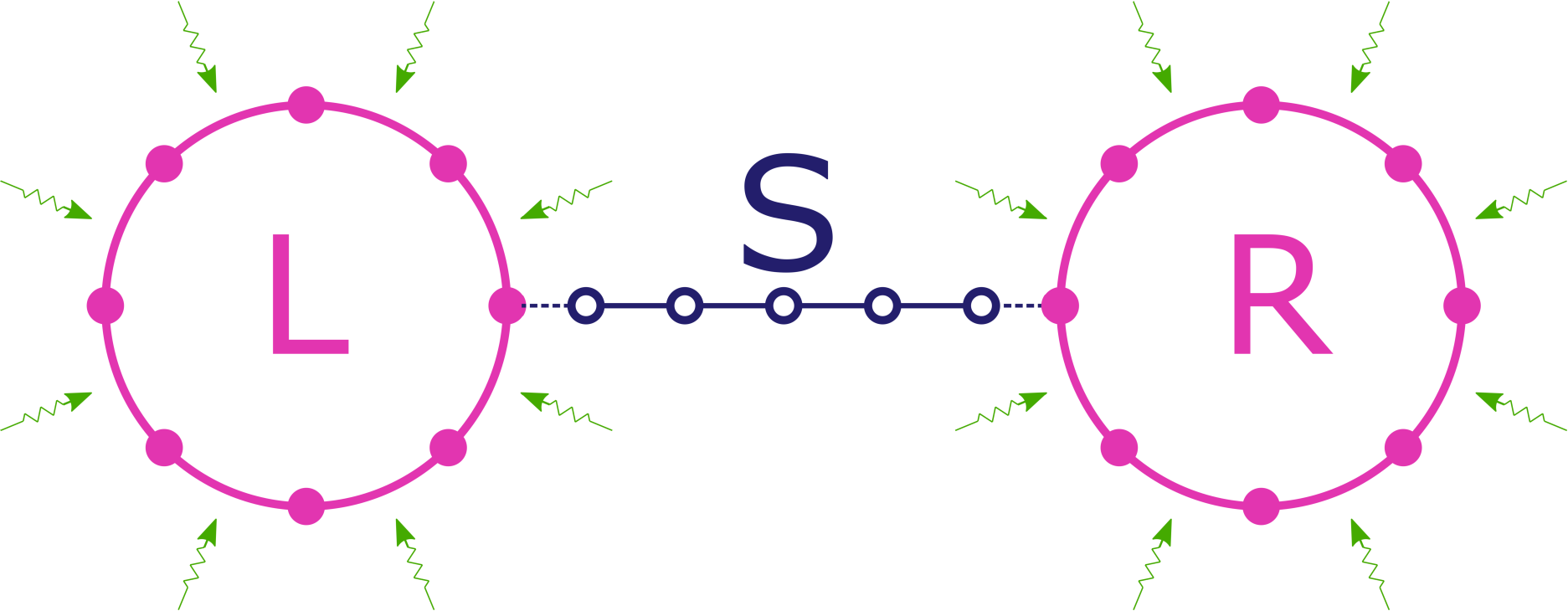}
\caption{The two-terminal setup where particle reservoirs/contacts are modelled by the tight-binding rings. }
\label{fig4_1}
\end{figure}

To specify the reservoir eigenstates we shall consider a tight-binding ring of the size $M$ where eventually $M\rightarrow\infty$. Thus, energies $E_k$ in the Fermi-Dirac distribution are $E_k=-J_r\cos(2\pi k/M)$. The lattice Hamiltonian has the standard form (\ref{tb_many}) where bosonic operators $\hat{a}_\ell$ and $\hat{a}_\ell^\dagger$ are substituted by the fermionic operators  $\hat{c}_\ell$ and $\hat{c}_\ell^\dagger$.  Let us now consider two reservoirs, each described by a ring and connected by a lattice which is attached to an arbitrary ring site $m$,  see Fig.~\ref{fig4_1}. With the reduced hopping parameter $\epsilon\ll J,J_r$, the interaction is described by
\begin{align} \label{coupling}
 \epsilon{\cal H}_{int}^{({\rm i})}=  \frac{\epsilon}{\sqrt{M}} \left( \hat{c}_{\rm i}^\dagger 
 \sum_{k=1}^{M} \hat{b}_k^{\rm (i)} \exp\left(i\frac{2\pi m}{M} k\right) + h.c. \right)   \;.
 \end{align}
Thus, our model consists of the lattice and two rings and the total density matrix $ {\cal R}_{tot}(t)$ obeys the master equation where the Lindblad operators act only on the ring quasimomentum states. Notice that all Hamiltonians and Lindblad operators are quadratic functions of the creation and annihilation operators. In this case we can obtain the master equation for the SPDM $\hat{\rho}_{tot}$ in closed form, 
\begin{align} \label{closed_1}
\frac{{d} \hat{\rho}_{tot}}{{d} t}=-i[\widehat{H}_{tot},\hat{\rho}_{tot}] +  \sum_{\rm i=L,R} {\cal L}_{\rm i}(\hat{\rho}_{tot})  \;, \\
\label{closed_2}
{\cal L}_{\rm i}[\hat{\rho}_{\rm i}^r(t)] = -\gamma[\hat{\rho}_{\rm i}^r(t) -\hat{\rho}_{\rm i}^0] \;, \quad 
\hat{\rho}_{\rm i}^0=\sum_k  \bar{n}_k |k\rangle\langle k | \;,
\end{align}
where $\hat{\rho}_{\rm i}^r(t)$ is the reduced SPDM of the respective reservoir and $\hat{\rho}_{\rm i}^0$ is its thermal state. The matrix $\hat{\rho}_{tot}(t)$  in Eq.~(\ref{closed_1}) is of dimension $(M+L+M)\times(M+L+M)$ and has block structure with nine blocks. Our prime interest is the central block which is the reduced density matrix $\hat{\rho}(t)$ of the carriers in the lattice.

It should be stressed that the introduced semi-microscopic model for the particle reservoir implicitly assumes the presence of a `super bath' because evolution of the total density matrix is not unitary. This assumption can be relaxed by introducing a fully-microscopic model for the particle reservoir. In the bosonic case such a model is discussed  in Ref.~\cite{preprint2} where the rings in Fig.~\ref{fig4_1} are given by the Bose-Hubbard Hamiltonians with  periodic boundary conditions.  Then, the time evolution of the total density matrix ${\cal R}_{tot}$ of the composed system is unitary. This, however, does not prevent relaxation of the ring SPDM to a thermal state Ref.~\cite{preprint2}.  Thus, Eq.~(\ref{closed_2}) can be justified also for microscopic models of a particle reservoir. 
\footnote{In this respect, i.e., in respect to microscopic models for particle reservoirs we mention the relevant theoretical studies Ref.~\cite{china1,china2} and the laboratory experiment Ref.~\cite{china3} with hard-core bosons, -- the system which is close to fermionic and spin systems, see chapter Ref.~\cite{chapter_18} in this volume.} Yet, there is one essential difference: by using a microscopic model we obtain $\gamma$ which depends on the model parameters like, for example, the reservoir temperature (i.e., the mean energy per particle), while in the semi-microscopic model $\gamma$ is an adjustable parameter. This is precisely what we need for the following discussion.

\subsection{Non-Markovian master equation} \label{sec4_4}
Having in hands the master equation for the total density matrix $\hat{\rho}_{tot}(t)$ we have two options of how to proceed further. The first option is to derive the master equation for the lattice density matrix $\hat{\rho}(t)$ by eliminating reservoirs. Using only a weak coupling (Born approximation) 
and considering the limit $M\rightarrow\infty$ this an integro-differential master equation was derived and analyzed in Ref.~\cite{129}. It was shown that this non-Markovian master equation does capture  the phenomenon of resonant transport, where the lattice conductance at zero temperature is given by a sum of Lorentzians
\footnote{We remind the reader that for zero temperature the chemical potential $\mu$ coincides with the Fermi energy $E_F$. Then the occupation numbers $\bar{n}_k$ in Eq.~(\ref{closed_2}) are  $\bar{n}_k=1$ if $|k|\le k_F$ and  $\bar{n}_k=0$ in the opposite case.}
\begin{align} \label{resonant}
\sigma(\mu) \approx \frac{1}{2\pi} \sum_{n=1}^L \frac{\alpha_n\epsilon^2}{2}\frac{(\gamma/2)}{(\gamma/2)^2+(\mu-E_n)^2} \;.
\end{align}
In this equation  $1/2\pi$ stands for the dimensionless conductance quantum, $E_n$ are eigenenergies of the tight-binding Hamiltonian (\ref{tb}), and we have the restriction $\epsilon^2/\gamma \ll \pi$ due to the Born approximation. 

The second option is to solve the master equation numerically for the total density matrix $\hat{\rho}_{tot}(t)$, whose central block gives us the lattice density matrix $\hat{\rho}(t)$. Notice that within this approach we may relax the Born approximation. However, the method has the disadvantage that one should run the solver several times for increasing ring size $M$ to check the convergence of the limit $M\rightarrow\infty$. In other words, for a given ring size $M$ we have to check that its further increase does not affect $\hat{\rho}(t)$. The empirical  rule states that convergence requires a critical value for $M$ that scales as $1/\gamma$. A hand-waving argument  behind this scaling is that non-zero $\gamma$ broadens the discrete energy levels of the finite-size reservoir proportionally to $\gamma$, which makes the reservoir spectrum effectively continuous as soon as the level spacing $E_{k+1}-E_k$  becomes smaller than $\gamma$.
\begin{figure}
\centering
\includegraphics[width=.5\textwidth]{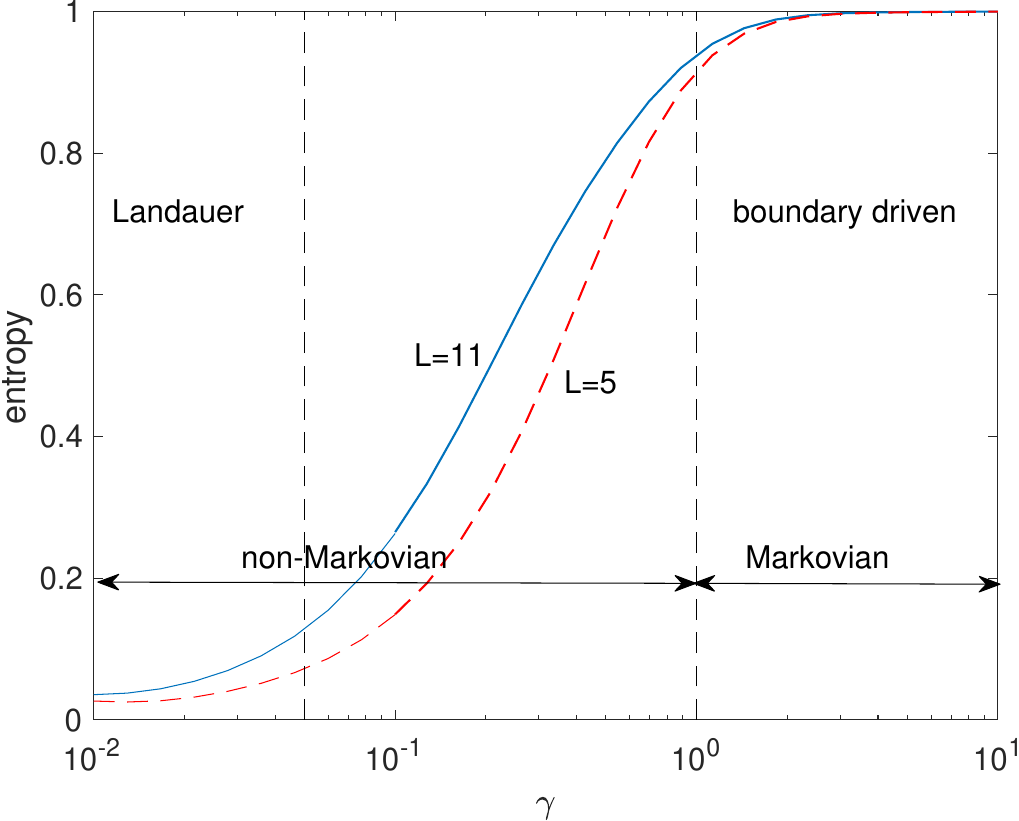}
\caption{The scaled entropy of the SPDM $\Delta\hat{\rho}$ of the fermionic carriers in the lattice as a function of the reservoir thermalization rate $\gamma$. Two limiting cases $S=0$ and $S=1$ correspond to the pure state and maximally mixed state, where the matrix $\Delta\hat{\rho}$ is proportional to the identity matrix.}
\label{fig4_2}
\end{figure}

\subsection{Landauer's {vs.} master equation approach} \label{sec4_5}
The key ingredient of the semi-microscopic  model for two-terminal transport is the finite relaxation rate $\gamma$ which we associate with the thermalization rate of quantum particles in reservoirs/contacts. As expected, in the limit of large $\gamma$ this model reproduces the results of the Markovian master equation (\ref{markovian_spdm}) with the particle exchange rate  $\tilde{\gamma}=\epsilon^2/\gamma$ \cite{129}. The opposite limit of small $\gamma$ relates the (non-Markovian) master equation approach to the Landauer theory. To show this we calculate the stationary density matrix $\bar{\rho}$ of the fermionic carriers in the lattice for $1/\beta=0$ and $\mu_{\rm i}=\mu \pm \Delta\mu/2$ and introduce the matrix $\Delta\hat{\rho}$ through the relation
\begin{align} 
\bar{\rho}=\bar{\rho}_0 +\Delta\mu \Delta\hat{\rho} \;,
\end{align}
where $\bar{\rho}_0$ is the stationary density matrix for $\Delta\mu=0$. Since the matrix $\bar{\rho}_0$ gives zero net current, the introduced matrix $\Delta\hat{\rho}$ determines the lattice conductance $\sigma(\mu)$. Figure \ref{fig4_2}  shows the von Neumann entropy of the matrix $\Delta\hat{\rho}$,
\begin{align} 
S=-{\rm Tr}[\Delta\hat{\rho}\ln \Delta\hat{\rho}] \;,
\end{align}
as a function of the reservoir thermalization rate $\gamma$. It is seen in Fig.~\ref{fig4_2} that in the limit $\gamma\rightarrow 0$ the entropy tends to zero and, thus, the matrix approaches some pure state $\Delta\hat{\rho}=|\Psi\rangle\langle \Psi |$, where $|\Psi\rangle=|\psi_1,\ldots,\psi_L \rangle$. Remarkably, the complex amplitudes $\psi_\ell$ in the latter equation coincide with the complex amplitudes which one obtains by solving the scattering problem for the plane wave with the energy given by the Fermi energy $E_F\equiv\mu$.  According to the common scattering theory the plane-wave transition probability  $|t(E)|^2$ for the lattice of the length $L$, which enters the Landauer formula for the lattice conductance, is given by a sum of $L$ Lorentzians,
\begin{align} \label{resonant_landauer}
| t(E) |^2 \approx \sum_{n=1}^L \frac{\Gamma_n^2}{\Gamma_n^2+(\mu-E_n)^2} \;, \quad  
\Gamma_n =\frac{\alpha_n \epsilon^2}{2} \;,\quad 
\alpha_n=| \langle \psi_n  | \ell=1\rangle\langle \ell=L | \psi_n\rangle  |  \;. 
\end{align}
It is easy to see that Eq.~(\ref{resonant_landauer}) and Eq.~(\ref{resonant}) are two limiting case  of the following equation \cite{131},
\begin{align} \label{landauer}
\sigma(\mu) = \frac{1}{2\pi}\sum_{n=1}^L \frac{\Gamma_n(\Gamma_n+\gamma/2)}{(\Gamma_n+\gamma/2)^2+(\mu-E_n)^2}  \;,  
\end{align}
which generalizes the Landauer formula onto the case of arbitrary $\gamma$. Indeed, in the limit  $\gamma\rightarrow 0$ Eq.~(\ref{landauer}) reproduces the celebrated Landauer result for the lattice conductance at zero temperature. Non-vanishing $\gamma$ broadens each resonant peak because of partial decoherence of the carrier states in the lattice which, in its turn, is due to dissipative dynamics of the carriers in the reservoirs attached to the lattice.  Then, in the case $\gamma\gg \Gamma_n$ we recover  Eq.~(\ref{resonant}) which was obtained by solving the non-Markovian master equation.

To conclude this section we notice that resonant transmission can be also observed for bosonic carriers \cite{123}. The necessary condition for this is that the width of Bose-Einstein distribution is smaller than the distances between the energies $E_n$, which is similar to the `fermionic condition'  that the width of the Fermi step at $E=E_F$ is smaller than the distances $|E_{n\pm1}-E_n|$. Going ahead, we also mention that in the bosonic case a particular interest is the resonant transport in the case where Bose particles are fully condensed in the left reservoir but the right reservoir is empty. One meets this situation in the quantum transport of microwave photons through the chain of coupled transmons 
\footnote{A transmon is a quantum nonlinear oscillator based on a Josephson junction. The presence of the junction makes microwave photons in a cavity effectively interacting and, thus, the chain of coupled transmons can be described by the Bose-Hubbard Hamiltonian.}, 
where the first transmon in the chain is excited by the microwave generator. We come back to this problem in Sec.~\ref{sec6}.

\section{Boundary driven Bose-Hubbard model}\label{sec5}
We proceed with the two-terminal transport of {\em interacting} Bose particles. In this section we shall analyze this problem within the framework of the Markovian master equation (\ref{markovian_many}) where $\widehat{\cal H}$ is the Hamiltonian  of the Bose-Hubbard model Eq.~(\ref{BH}).  As was stated in the previous section, in the case $U=0$ and $\bar{n}_{\rm L}\ne \bar{n}_{\rm R}$ we have a stationary current Eq.~(\ref{current}). It is also easy to show that stationary populations of the lattice sites, i.e., the diagonal elements of the steady-state SPDM are  $\rho_{\ell,\ell}=(\bar{n}_{\rm L}+\bar{n}_{\rm R})/2$,  except the first and the last sites for which we have a slightly different equation. We address the question of how the inter-particle interaction modifies the stationary current and stationary populations of the lattice sites.

\subsection{Ballistic {\em vs.} diffusive currents} \label{sec5_1}
The simplest approach to the posed problem is based on the conjecture of Sec.~\ref{sec3_5}  that a weak inter-particle interaction can be taken into account by including the local dephasing operator Eq.~(\ref{decoherence}) in the master equation for the SPDM,  
\begin{align} \label{boundary}
\frac{d \hat{\rho}}{d t}=-i[\widehat{H},\hat{\rho}] +\sum_{\rm i=L,R}{\cal L}_{\rm i}(\hat{\rho}) + \widetilde{\cal L}(\hat{\rho}) \;.
\end{align}
 Luckily, the stationary solution of the displayed master equation is known analytically \cite{Znid10b},
\begin{equation}\label{diffusive}
\bar{j}=\frac{J^2\gamma}{J^2+\gamma^2+\eta\gamma (L-1)} \frac{\bar{n}_{\rm L}-\bar{n}_{\rm R}}{2} \;,
\end{equation}
where we changed notation $\tilde{\gamma} \rightarrow \gamma$. Comparing this equation with Eq.~(\ref{current}) we see that the stationary current now depends on the lattice length $L$  and asymptotically $\bar{j} \sim 1/L$. In the physical literature, this change in the expression for the stationary current is referred to as the transition from the {\em ballistic} transport regime, where the current is independent of $L$, to the {\em diffusive} transport regime, where the current is inverse proportional to $L$.
\footnote{ Notice that this definition of the ballistic and diffusive currents is asymptotic in the thermodynamic limit $L\rightarrow \infty$. For finite lattices discrimination between ballistic and diffusive transport regimes is not so obvious \cite{136}.}
Another prediction of the analytical study of Eq.~(\ref{boundary}) is that stationary populations of the lattice cites form a straight line with the slope $(\bar{n}_{\rm L}-\bar{n}_{\rm R})/L$. It should be remembered, however, that a microscopic justification of the  master equation (\ref{boundary}) is absent and, thus, all results obtained on the basis of this equation can be questioned. In what follows we shall analyze the stationary current of interacting bosons by using a more rigorous approach of the Truncated Wigner Function (a variant of the pseudoclassical approach for open systems) but first, to introduce this technique to the reader, we consider the open one-site Bose-Hubbard model.

\subsection{Pseudoclassical analysis of the open one-site Bose-Hubbard model} \label{sec5_2}
\begin{figure}
\centering
\includegraphics[width=.5\textwidth]{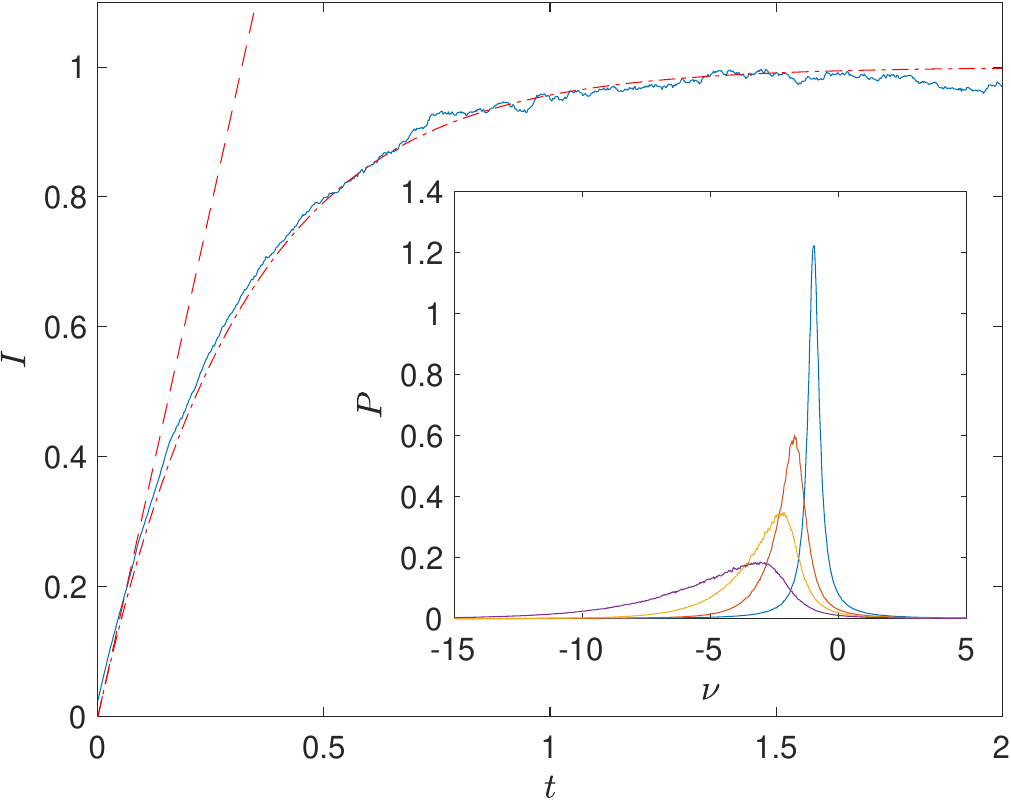}
\caption{Dash-dotted line: Dynamics of $I(t)=N(t)/\bar{n}$ according to the master equation (\ref{oscillator2}). The parameters are $\gamma=0.5$ and $\bar{n}=10$. The initial condition is the oscillator ground state $| 0\rangle$. Dashed line: Solution of the master equation for $D=0.5$ and $\gamma=0$. Blue solid line: Solution of the Langevin equation (\ref{langevin}) averaged over 4000 realizations of the random process $\xi(t)$.  The inset shows the oscillator spectral density $P(\nu)$ for $\Omega=1$ and different values of nonlinearity $g=0,0.5,1,2$, from top to bottom. The figure is borrowed from Ref.~\cite{116}.}
\label{fig5_1}
\end{figure}

Let us consider the one-site Bose-Hubbard model or, equaivalently, the quantum nonlinear oscillator
%
which is coupled to a reservoir of Bose particles,
\begin{align} \label{oscillator2}
\frac{d {\cal R}}{d t}=-i[\widehat{\cal H},{\cal R}] +{\cal L}({\cal R}) \;, \quad
\end{align}
In Eq.~(\ref{oscillator2}) $\widehat{\cal H}$ is the one-site Bose-Hubbard Hamiltonian and  ${\cal L}({\cal R})$ is the Lindblad operator given by the sum of the gain and loss operators, 
\begin{align} 
{\cal L}_{loss}({\cal R})=-\frac{\gamma}{2} (\bar{n}+1)
\left(\hat{a}^{\dagger} \hat{a}{\cal R }-2\hat{a}{\cal R }\hat{a}^{\dagger} +{\cal R }\hat{a}^{\dagger}\hat{a} \right)  \;,\quad
{\cal L}_{gain}({\cal R})= -\frac{\gamma}{2}\bar{n}
\left(\hat{a}\hat{a}^{\dagger}{\cal R }-2\hat{a}^{\dagger}{\cal R }\hat{a} +{\cal R }\hat{a}\hat{a}^{\dagger} \right)  \;, 
\end{align}
which are parametrized by  the reservoir particle density $\bar{n}$. It is instructive to rewrite the sum of the loss and gain operators as a sum of the other two operators, 
\begin{align} \label{lindblad3}
{\cal D}({\cal R})=-\frac{D}{2}\left( [\hat{a},[\hat{a}^\dagger,{\cal R}]] + [\hat{a}^\dagger,[\hat{a},{\cal R}]] \right)  \;, \quad
{\cal G}({\cal R})=-\frac{\gamma}{2}(\hat{a}^\dagger\hat{a}{\cal R}-2\hat{a}{\cal R}\hat{a}^\dagger + {\cal R}\hat{a}^\dagger\hat{a}) \;,
\end{align}
where $D=\gamma\bar{n}$. As it will be clear in a moment, the operator ${\cal D}({\cal R})$ is a diffusion term with diffusion coefficient $D$ and the operator ${\cal G}({\cal R})$ is a friction term with friction coefficient $\gamma$. Indeed, considering $D$ and $\gamma$ to be independent parameters we  find that for $D=0$ but $\gamma\ne0$ the nonlinear oscillator  relaxes to the ground state, while in the case $D\ne0$ but $\gamma=0$ the mean particle number  $N(t)={\rm Tr}[\hat{n} {\cal R}(t)]$ grows linearly in time. In the general case $D\ne0$ and $\gamma\ne0$ the quantity $N(t)$ exponentially converges to the value $D/\gamma=\bar{n}$, see Fig.\ref{fig5_1}.

Now we shall obtain the same results by using the truncated Wigner function approximation. The first step is to rewrite the master equation (\ref{oscillator2}) in terms of the Wigner function (\ref{wigner}). Using the Wigner-Weyl transformation we have after some algebra
\begin{align} \label{}
\frac{\partial f_{qu}}{\partial t}= \{ H,f_{qu} \} + {\cal D}(f_{qu})+{\cal G}(f_{qu}) + O\left(\frac{1}{\bar{n}}\right) \;,
\end{align}
where 
\begin{align} \label{}
{\cal D}(f_{qu})= D\frac{\partial^2 f_{qu}}{\partial a \partial a^*}  \;, \quad 
{\cal G}(f_{qu})=\frac{\gamma}{2}\left( \frac{\partial (a f_{qu})}{\partial a} + \frac{\partial (a^*f_{qu})}{\partial a^*} \right) \;.
\end{align}
%
Next, we drop all terms which are proportional to a power of the effective Planck constant $\hbar'=1/\bar{n}$.  This converts the equation for the Wigner function into the Fokker-Planck equation for a classical distribution function. Finally, we map this  Fokker-Planck equation to the Langevin equation,
\begin{equation}\label{langevin}
i\dot{a}= \frac{\partial H}{\partial a^*} - i\frac{\gamma}{2} a + \sqrt{\frac{D}{2}} \xi(t)  \;,  \quad
H=\Omega a^*a+\frac{g}{2}(a^*a)^2 \;,
\end{equation}
where $\xi(t)={\rm Re}[\xi(t)]+i {\rm Im}[\xi(t)]$ is $\delta$-correlated noise with standard deviation equal to unity for both the real and imaginary parts.  Thus,  we reduced the quantum problem to dynamics of the damped classical oscillator which is subject to the white noise. The quantity $I(t)=\langle\overline{a^*(t)a(t)}\rangle$, where the over-line denotes the average over different realizations of the Wiener process $\xi(t)$  and the angular brackets the average over quantum ensemble of initial conditions, is depicted in Fig.~\ref{fig5_1} by the solid line. A nice agreement with the exact result is noticed.

Concluding this subsection we mention that the described pseudoclassical approach allows us to introduce an important characteristic of the system -- the oscillator spectral density $P(\nu)$,
\begin{equation} 
P(\nu)=\overline{|a(\nu)|^2} \;, \quad a(\nu)=\lim_{T\rightarrow\infty} \frac{1}{T} \int_0^T a(t) e^{i\nu t} {\rm d}t \;.
\end{equation}
Here the limit $T\rightarrow\infty$ insures that $P(\nu)$ refers to the stationary regime. (In practice this requires the evolution time $T$ to be larger than the relaxation time $T_\gamma \sim 1/\gamma$.) Although the spectral density $P(\nu)$ has no a quantum analog, it is quite useful for distinguishing between different transport regimes in the quantum boundary driven Bose-Hubbard model.

\subsection{Stationary current across the Bose-Hubbard chain} \label{sec5_3}
In this subsection we calculate the current between two reservoirs of Bose particles connected by the Bose-Hubbard chain by using the pseudoclassical method which was introduced in the previous subsection. As already mentioned in Sec.~\ref{sec3_1}, from the viewpoint of classical mechanics the Bose-Hubbard model is a chain of $L$ coupled nonlinear oscillators. And, as was shown in the previous subsection, the effect of reservoirs is that  the first and last oscillator in the chain are subject to friction with coefficient $\gamma$ and, simultaneously, they are excited  by a stochastic force of magnitude $\sqrt{(D/2)(\bar{n}_{\rm i}/\bar{n}_{\rm L})}$. 
\footnote{To be specific we define the effective Planck constant as $\hbar'=1/\bar{n}_{\rm L}$.}
Notice that due $\bar{n}_{\rm L} \ne \bar{n}_{\rm R}$ the force magnitude is different for the edge oscillators.  The most intuitive case is $\bar{n}_{\rm R}\approx 0$ where the last oscillator is just damped.  Then the particle current across the chain is associated with the transport of excitations from the first to the last oscillator.
\begin{figure}
\centering
\includegraphics[width=.7\textwidth]{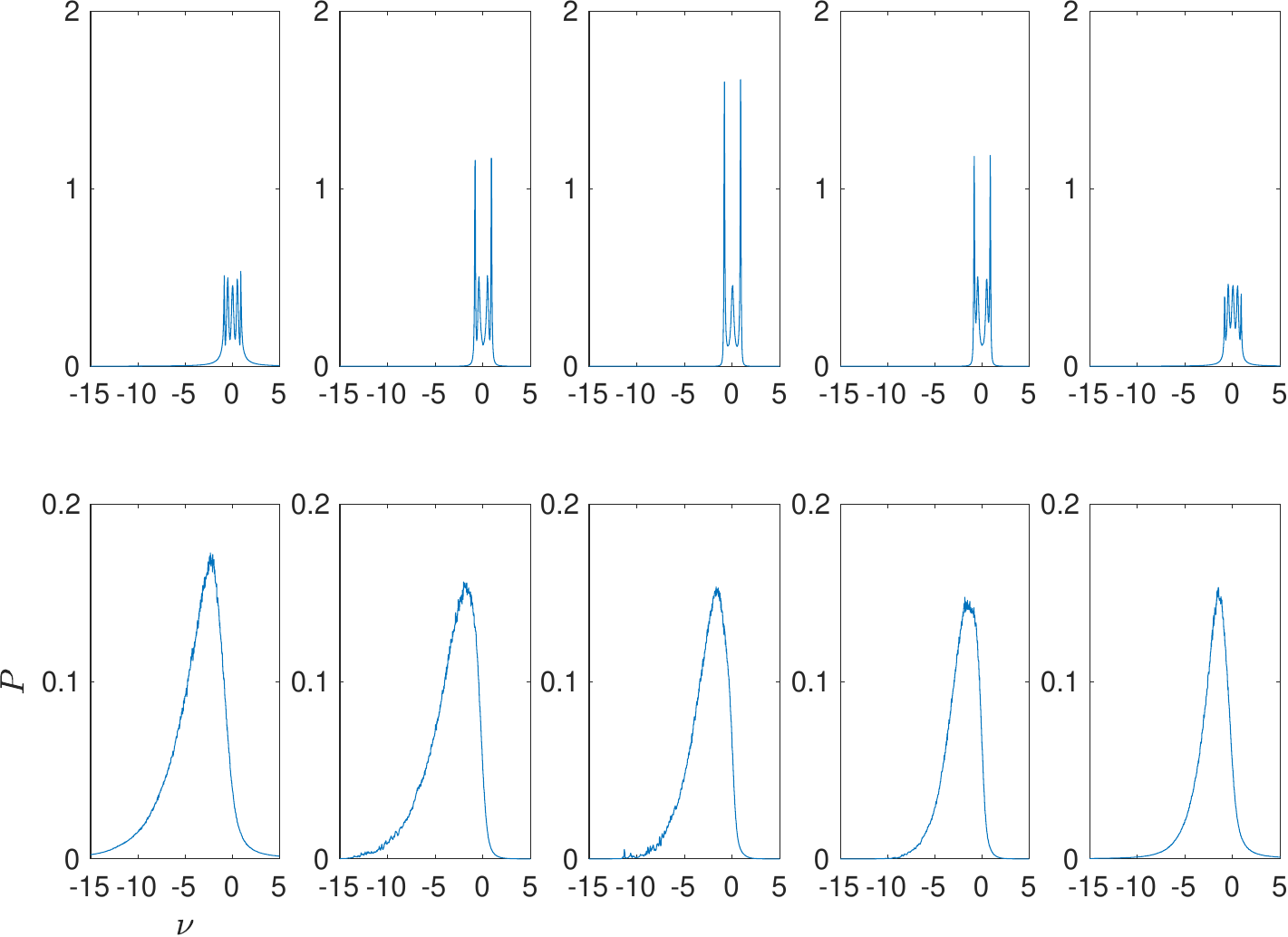}
\caption{Oscillators spectral densities for $g=0$, upper row, and $g=2$, lower row. The other parameters are $J=1$, $\gamma=0.5$, $D_{\rm L}=0.5$, and $D_{\rm R}=0.25$, corresponding to $\bar{n}_{\rm R}=\bar{n}_{\rm L}/2$. The figure is borrowed from Ref.~\cite{116}.}
\label{fig5_2}
\end{figure} 
\begin{figure}
\centering
\includegraphics[width=.7\textwidth]{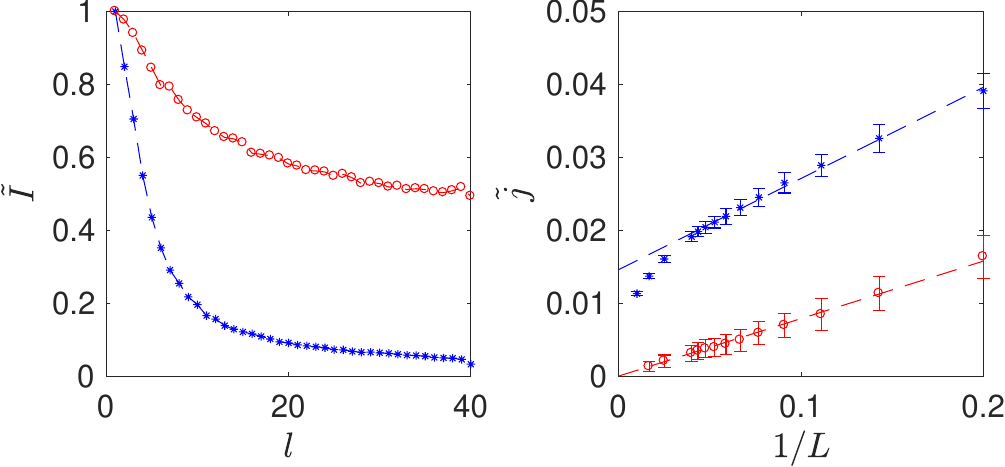}
\caption{Left panel: Normalized diagonal elements of the stationary single-particle density matrix for $L=40$, $D_{\rm R}=D_{\rm L}/2$ (open circles) and $D_{\rm R}=0$ (asterisks). Right panel: The stationary current as the function of the inverse chain length for the two considered cases. The bars show statistical error due to finite number of realizations of the stochastic process. The figure is borrowed from Ref.~\cite{116}.}
\label{fig5_3}
\end{figure}

First, we analyze the transition from the ballistic to the diffusive  regime for short lattices, i.e., without considering the limit $L\rightarrow\infty$. Within the framework of the pseudoclassical approach this transition is well seen in the spectral densities of the oscillators. For $L=5$ these densities are depicted in Fig.~\ref{fig5_2} for $g=0$, upper row, and $g=2$, lower row. It is seen in Fig.~\ref{fig5_2} that for vanishing inter-particle interactions the stochastic driving of the edge oscillators excites delocalized modes of the system, whose eigenfrequencies  are approximately given by $\omega_k=J\cos(2\pi k/L)$. This insures efficient transport of excitations from the first to the last oscillator, which quantum-mechanically corresponds to ballistic current. This situation is opposed to the diffusive regime for $g=2$ where one finds fundamentally different spectral densities. 

Next, we discuss long lattices up to $L=100$. 
The left panel in  Fig.~\ref{fig5_3} shows the diagonal elements of the single-particle density matrix calculated for $L=40$, $g=2$, $\gamma=0.5$, $D_{\rm L}=0.5$, and $D_{\rm R}=0.25$ (red open circles) and $D_{\rm R}=0$ (blue asterisks). It is seen in Fig.~\ref{fig5_3}(a) that populations of the lattice sites strongly deviate from the prediction of the phenomenological master equation (\ref{boundary}). Thus, Eq.~(\ref{boundary}) is oversimplified and it does not   predict correct results in the general case of arbitrary  $\bar{n}_{\rm L}$ and $\bar{n}_{\rm R}$. We also mention that in this general  case the functional dependence of the stationary current on the system size is given by $\bar{j} \sim L^{-\alpha}$ where the exponent $\alpha$ depends on the ratio $\bar{n}_{\rm R}/\bar{n}_{\rm L}$.  For example, for  $\bar{n}_{\rm R}/\bar{n}_{\rm L}=0,1/2,3/4$ the exponent is $\alpha=0.67,0.93,0.95$, respectively. Thus, strictly speaking, the transport regime is subdiffusive, approaching the diffusive regime ($\alpha=1$) only in the limit  $\bar{n}_{\rm R}\rightarrow\bar{n}_{\rm L}$.

\subsection{Spectral statistics of the stationary density matrix} \label{sec5_4}
The approximate methods discussed  above allow us to calculate the SPDM $\hat{\rho}(t)$ but not the many-body density matrix ${\cal R}(t)$ which contains information about quantum correlations in the system. Since the closed/conservative Bose-Hubbard model is chaotic,  we may expect the stationary many-body density matrix of the open Bose-Hubbard to also be chaotic.  Unfortunately, when checking this hypothesis we encounter two problems which prevent us from a definite conclusion: (i) it is very hard to calculate the stationary many-body density matrix for a reasonable system size; (ii) since the matrix  ${\cal R}(t)$ is formally infinite, it is not clear of how truncation parameter affects the results and what would be the appropriate value for this parameter. Fortunately, we can get rid of these problems by considering two-terminal quantum transport with a fixed number of particles. In this subsection we introduce such a model. 
\begin{figure}[b]
\centering
\includegraphics[width=10.0cm,clip]{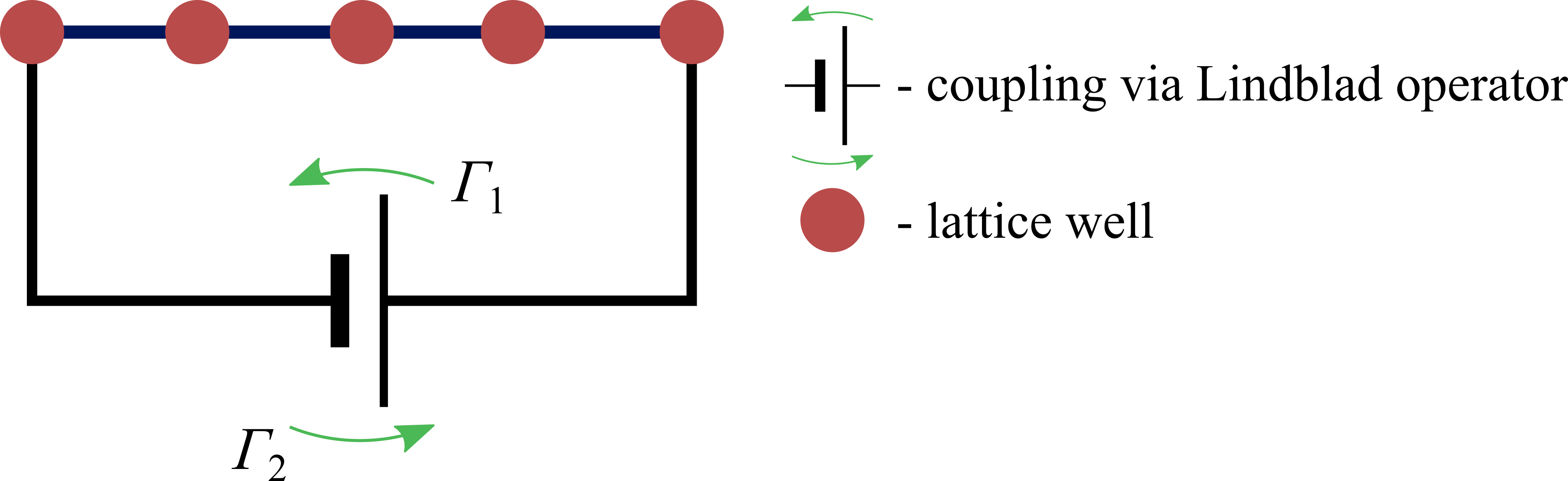}
\caption{Pictorial presentation of the model (\ref{fixed}).}
\label{fig5_6}
\end{figure}

Following Ref.~\cite{130} we consider the Bose-Hubbard chain of length $L$ with incoherent coupling of its first and last sites. Namely, the coupling is described by the the following Lindblad operators,
\begin{equation}
{\cal L}_1({\cal R})=-\Gamma_1\left(\hat{V}^{\dagger}\hat{V}{\cal R} 
  - 2\hat{V}{\cal R}\hat{V}^{\dagger}  + {\cal R}\hat{V}^{\dagger}\hat{V} \right) 
   \;, \quad
{\cal L}_2({\cal R})= -\Gamma_2\left(\hat{V}\hat{V}^{\dagger}{\cal R} 
 - 2\hat{V}^{\dagger}{\cal R}\hat{V} + {\cal R}\hat{V}\hat{V}^{\dagger} \right)  \;,
\end{equation}
where $\hat{V}=\hat{a}_1^\dagger \hat{a}_L$.  Thus, the master equation for the carrier density matrix ${\cal R}$ reads
\begin{equation}
\label{fixed}
\frac{\partial {\cal R}}{\partial t}=-i[\widehat{\cal H}, {\cal R}] +{\cal L}_1({\cal R}) + {\cal L}_2({\cal R}) \;,
\end{equation}
where $\widehat{\cal H}$ is the Bose-Hubbard Hamiltonian. The Lindblad operator ${\cal L}_1({\cal R})$ in Eq.~(\ref{fixed}) induces  incoherent transport of carriers from the last to the first sites, while the operator ${\cal L}_2({\cal R})$ is responsible for incoherent transport in the reverse direction. If the rates $\Gamma_1 \ne \Gamma_2$, there is a non-zero current in the clockwise or counterclockwise direction depending on the inequality relationship between these two relaxation constants. We notice that, by an analogy with electronic devices, the introduced Lindblad operators mimic a battery which induces stationary currents in electric circuits, see Fig.~\ref{fig5_6}.

\begin{figure}
\centering
\includegraphics[width=9.0cm,clip]{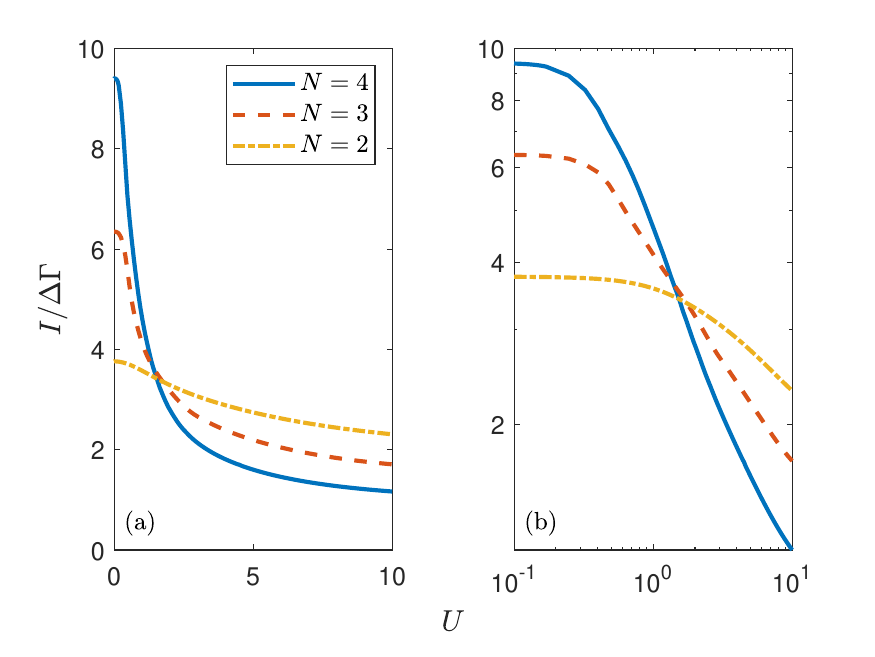}
\caption{The conductance of the  6-site Bose-Hubbard chain as a function of the interaction constant $U$ for $\Gamma=0.04$ and three different values of the particle number $N=2,3,4$ in the linear (a) and logarithmic (b) scales. The figure is borrowed from Ref.~\cite{130}.}
\label{fig5_5}
\end{figure}

It is easy to prove that if $\Gamma_1 = \Gamma_2$ the steady-state density matrix is proportional to the identity matrix, namely, ${\cal R}=\widehat{1}/{\cal N}$, where ${\cal N}$ is the dimension of the Hilbert space. Focusing on the liner response regime where $\Gamma_1=\Gamma +\Delta\Gamma/2$ and $\Gamma_2=\Gamma -\Delta\Gamma/2$ we have 
\begin{equation}
\label{b1}
{\cal R}=\frac{\widehat{1}}{{\cal N} }+\Delta\Gamma \Delta{\cal R} \;, 
\end{equation}
where the matrix $\Delta{\cal R}$ determines the system conductance. Next, we diagonalize the matrix  $\Delta{\cal R}$ and calculate its spectrum. It was shown in Ref.~\cite{130} that with an increasing interaction constant $U$ the level-spacing distribution rapidly changes from Poisson to GUE Wigner-Dyson distributions and then back to Poisson distribution. This change in the spectral statistics is well reflected in the magnitude of the stationary current across the Bose-Hubbard chain, see Fig.~\ref{fig5_5}. The rapid decrease of the current in the interval $0\le U<1$ is correlated with the transition to chaos in the closed Bose-Hubbard system.  This rapid decrease is followed by a slow decrease, which is correlated with the transition back to Poisson statistics and attributed  to fermionization of Bose particles. Indeed, for $U\rightarrow\infty$ weakly interacting bosons become hard-core bosons, which is an integrable system similar to spinless fermions. We stress that the limit of large $U$ is a pure quantum transport regime where the pseudoclassical approach cannot be applied in principle. 


\subsection{Intermediate conclusion} \label{sec5_5}
We studied the stationary current of interacting Bose particles across the tight-binding chain of the finite length $L$ by using three different methods: (i) the phenomenological master equation for the SPDM of weakly interacting bosons $\hat{\rho}(t)$; (ii) the pseudoclassical approach which allows us to calculated this matrix by mapping the quantum problem into a classical one; (iii) the exact numerical analysis of the stationary many-body density matrix ${\cal R}$ of $N$ interacting bosons. These three approaches give consistent results and lead to the following main conclusion. With an increase in the interaction strength the ballistic transport regime changes to diffusive transport and this crossover is correlated with transition to chaos in the closed/conservative Bose-Hubbard model.

\section{Many-body dissipative systems with periodic driving} \label{sec6}
In the previous section we analyzed the boundary driven Bose-Hubbard model by using the Markovian master equation. This equation is valid for a high temperature in the reservoirs, where the Bose-Einstein distribution is almost flat, and large values of the thermalization rate $\gamma$. In the pseudoclassical approach this situation corresponds to excitations of the edge oscillators by a broad-band noise where the band width is larger than the width $2J$ of the conductance band of the lattice. In the opposite limit of low temperatures and small values of the thermalization rate one has to use the non-Markovian master equation. In the pseudoclassical approach  this corresponds to the excitation of the edge oscillators by a narrow-band noise  with some characteristic frequency $\omega$. The limiting case of narrow-band noise leads to periodic driving of the system, where the first oscillator is excited by a monochromatic force with the frequency $\omega$ while the last oscillator is usually considered to be purely dissipative.  One meets this situation, for example, in experiments with super-conducting circuits where the first transmon in the chain of coupled transmons 
is excited by the microwave generator and the transmitted signal is measured by the vector analyzer attached to the last transmon \cite{Fitz17,Fedo21}.  In other words,  microwave photons are injected into the first site of the Bose-Hubbard chain and withdrawn  from the last site. An important   advantage of super-conducting circuits as compared to cold Bose atoms in optical lattices is that experimentalists can easily realize quasi one-dimensional chains like, for example, the diamond lattice or the flux rhombic lattice.
\footnote{We also mention a similar experiment with photonic crystals \cite{Cace22}.}

\subsection{Driven dissipative nonlinear oscillator} \label{sec6_1}
Similar to the previous section we begin with the dissipative one-site Bose-Hubbard model which is now driven by an external periodic force with the frequency $\omega$,          
\begin{equation} \label{oscillator3}
\widehat{\cal H}=E_0 \hat{a}^\dag\hat{a}  + \frac{U}{2} \hat{n}(\hat{n}-1) +\varepsilon\left(\hat{a}^\dagger e^{-i\omega t} + \hat{a} e^{i\omega t} \right) \;.
\end{equation}
The governing master equation has the standard form (\ref{oscillator2}) with the Lindblad operator ${\cal L}({\cal R})={\cal G}({\cal R})$,  Eq.~(\ref{lindblad3}). The classical counterpart of this system is the driven dissipative nonlinear oscillator,
\begin{equation} \label{bistability}
i\dot{a}=\left(\Omega-i\frac{\gamma}{2}\right) a+ g|a |^2 a + \varepsilon e^{-i\omega t}  \;,
\end{equation}
which is known to be the simplest dynamical system showing bistability.  Namely, Eq.~(\ref{bistability}) has two steady-state solutions which correspond to two limit cycles with different oscillation amplitudes \cite{Land76}. However, the pioneering studies of the quantum driven dissipative oscillator by Drummond and Walls \cite{Drum80} showed that the solution of the governing master equation is unique, which questioned the principle of the quantum-classical correspondence. This contradiction was resolved by noticing that  the quantum oscillator may also relax to two different states but only one of them is the true steady state while the other one is a metastable state. Which of two quantum limit cycles is a metastable state depends on the value of a control parameter (for example, the detuning $\delta=\Omega-\omega$) and there can also be situations where the steady-state density matrix is a `superposition' of two limit cycles.
\footnote{A brief tutorial review of the driven dissipative quantum nonlinear oscillator can be found in Ref.~\cite{my_preprint}.}

The problem  becomes much more complicated when we consider  the $L$-site Bose-Hubbard model. The driven dissipative system of $L$ coupled  nonlinear oscillators can simultaneously (i.e., for a fixed set of parameters) have many attractors of different types -- simple focuses, limit cycles, periodic attractors, and even chaotic attractors -- and these attractors typically show a rather complex bifurcation diagram \cite{Gira20}.  Because of this it is hard to predict which of the many attractors corresponds to the true steady state in the quantum case.  Fortunately, the pseudoclassical approach is helpful even in this situation if we include the first order quantum corrections to the classical equation of motion of the open Bose-Hubbard model. The procedure goes as follows.

First, one introduces the dimensionless  effective Planck constant $\hbar'$ which determines 
how close the quantum system is to its classical counterpart. On the formal level this effective Planck constant defines the scaling law for the system parameters which keeps  the classical dynamics invariant. For example, for the system (\ref{oscillator3}) this scaling law results in the following quantum Hamiltonian
\begin{equation} 
\widehat{H}=\hbar' \left(\Omega -\frac{\hbar' g}{2} \right) \hat{a}^\dag\hat{a}  
+ \frac{\hbar'^2 g}{2} \hat{a}^\dagger\hat{a}^\dagger \hat{a}\hat{a} 
+\sqrt{\hbar'}\varepsilon  \left(\hat{a}^\dagger e^{-i\omega t} + \hat{a} e^{i\omega t} \right) \;.
\end{equation}
One also has to include the effective Planck constant in the master equation, i.e.,  $-i[\widehat{\cal H},{\cal R}] \rightarrow -i/\hbar'[\widehat{\cal H},{\cal R}]$ and $\gamma \rightarrow \hbar'\gamma$. Physically,  this scaling means that the classical limit is reached if we decrease the microscopic interaction constant $U$ proportionally to $\hbar'$ and simultaneously increase the magnitude of the driving force proportionally to $1/\sqrt{\hbar'}$. In the next step we derive the equation  for the Wigner function where we neglect the terms $O(\hbar'^2)$ but keep the term proportional to $\hbar'$. This term appears to be the diffusive term $D\partial^2f_{qu}/\partial a\partial a^*$ with $D=\hbar'\gamma$. Thus, when implementing  Monte-Carlo simulations, the classical Eq.~(\ref{bistability}) should be complemented by a stochastic term $\sqrt{\hbar'\gamma/2}\xi(t)$.  This seemingly minor modification of the classical equation of motion has the fundamental consequence that the stationary solution for the classical distribution function $f_{cl}=f(a,a^*; t)$ is unique and, thus, we can use it as the first order approximation for the quantum distribution function $f_{qu}=f(a,a^*; t)$.  We also mention in passing, that for vanishing driving $\epsilon=0$, the irreducible quantum noise of the intensity $\sqrt{\hbar'\gamma/2}$ does not allow the damped classical oscillator to go below the ground state energy $\hbar'\Omega/2$ of the quantum oscillator.

\subsection{Conduction transition in the flux rhombic lattice} \label{sec6_3}
To further highlight the power of the pseudoclassical  approach we consider the conduction transition in the flux rhombic lattice. This system is of vivid theoretical and experimental interest due to the phenomenon of  Aharonov-Bohm caging \cite{Cace22,Mart23,128}. The rhombic lattice is constructed by connecting $L$ rhombs, whose vertices/sites we shall label by $A,B,D,C$, into a chain where the $D$ site of the first rhomb is identified with the $A$ site of the second rhomb and so on. Of particular interest is the case where the hopping matrix elements between $A$ and $B$ sites and $A$ and $C$ sites have opposite signs. Then, because of complete destructive interference, the quantum particle cannot reach the $D$ site and the rhombic lattice is insulating independent of the magnitude of the driving force $\epsilon$. This conclusion, however, is valid only for non-interacting particles and for a finite $g$ there is a critical value $\epsilon_{cr}$ above which the lattice becomes conducting. We want to find this critical strength of driving.
\begin{figure}
\centering
\includegraphics[width=12.cm]{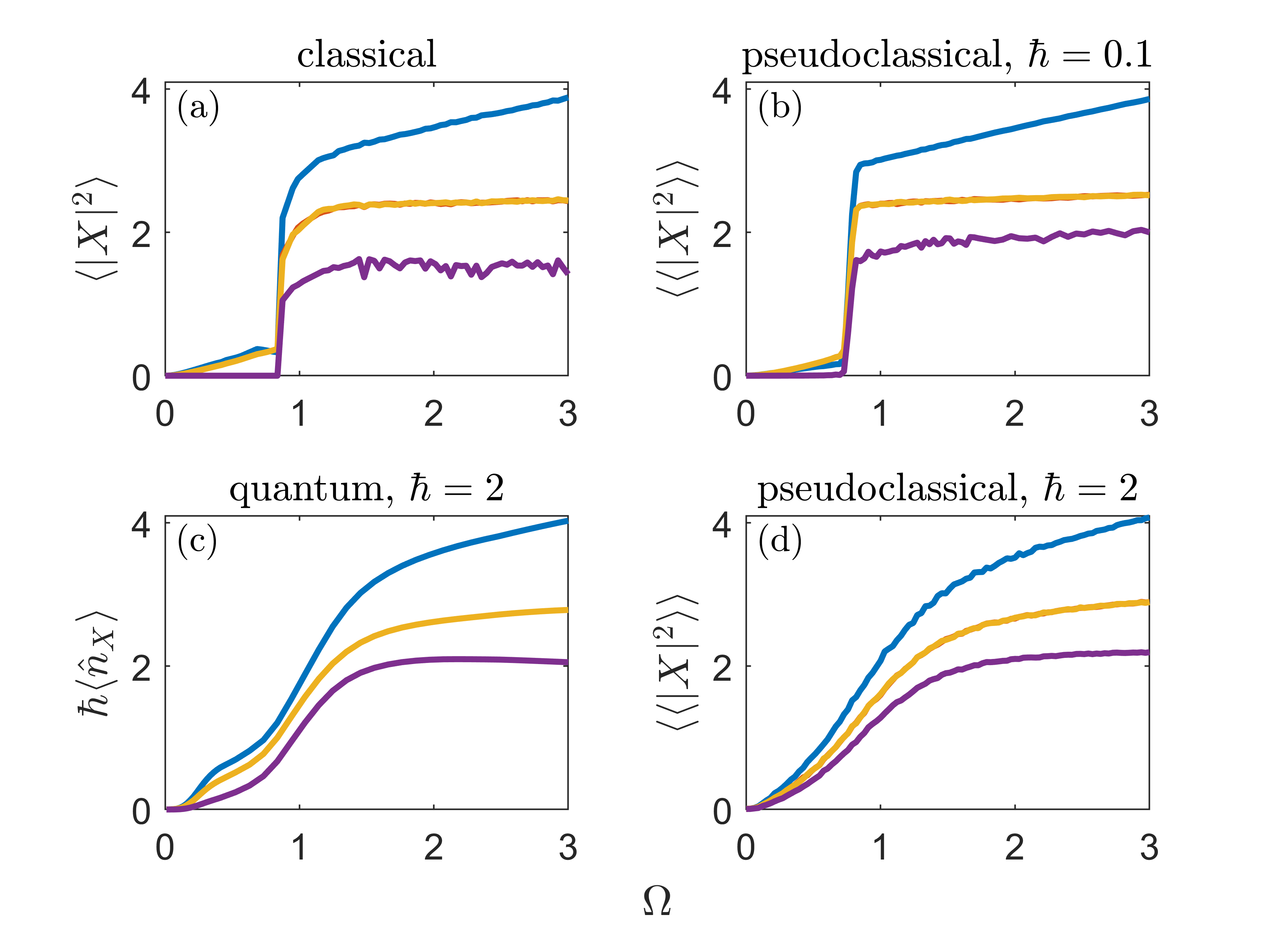}
\caption{Left column: (a)The mean squared amplitudes of the classical oscillators in the stationary regime versus $\varepsilon$, averaged over random initial conditions; (c) The mean populations of the rhomb sites for $\hbar'=2$ in the quantum case versus $\varepsilon$.
Right column:  The mean squared population of the rhomb sites for the pseudoclassical approach, (b) $\hbar'=0.1$ and (d) $\hbar'=2$. For all cases $|C|^2$ - blue curves (top), $|A|^2$ and $|B|^2$ - yellow curves (middle), $|D|^2$ - purple curves (bottom). The other  parameters are $\Omega-\omega=-0.5$, $g=0.5$,  and $\gamma=0.2$. The figure is borrowed from Ref.~\cite{128}.}
\label{fig6_1}
\end{figure}

Following the setting of the laboratory experiment \cite{Mart23} we consider single rhomb and simulate classical dynamics for the experimentally relevant initial conditions where at $t=0$ all four sites  are almost empty. The system parameters are $g=0.5$, $\gamma=0.2$, $\Omega-\omega=-0.5$, and we choose $\varepsilon$ to be our control parameter. We remark that for $\gamma=0$ and  $\varepsilon>\varepsilon_{cr}$ the system enters a chaotic regime where the complex amplitude of four nonlinear oscillators, which we denote by $X(t)$, $X=A,B,D,C$, are projections of the chaotic trajectory in the eight-dimensional phase space of the system \cite{128}. In the dissipative case $\gamma\ne0$, however, we have to analyze the bifurcation diagram of several attractors whose basins include the vicinity of the point $|X|=0$. To be specific, we consider $|X|=0.001$ with uniformly distributed random phases. The stationary populations of the rhomb sites $\langle |X(t)|^2\rangle$, where the angular brackets denote the averaged over the ensemble of initial conditions, are shown in Fig.~\ref{fig6_1}(a)  as the function of $\epsilon$. The point to draw the reader's attention to is a jump in the population of the $D$ site at $\varepsilon=\varepsilon_{cr}\approx 0.9$, which indicates the breakdown of the Aharonov-Bohm caging.

Next, we consider numerical simulations of the quantum dynamics which  are feasible only for $\hbar'\ge1$. Stationary populations of the rhomb sites in the quantum case are shown in Fig.~\ref{fig6_1}(c). Comparing Fig.~\ref{fig6_1}(c) with Fig.~\ref{fig6_1}(a) we conclude that quantum fluctuations smooth the sharp transition between the insulating and conducting phases. Remarkably, the pseudoclassical approach, where we additionally average the quantity $\langle |X(t)|^2\rangle$ over different realizations of the irreducible quantum noise, fairly reproduces the quantum result,  see Fig.~\ref{fig6_1}(d), as well as the convergence to the classical case,  see Fig.~\ref{fig6_1}(b).

\section{Conclusion} \label{sec7}
Let us briefly summarize the reviewed problem of chaotic transport in quantum systems, highlighting the main results and ideas. 

In Sec.~\ref{sec2} we considered chaotic dynamics in single-particle systems. The main message of this section is that there is no need to seek chaotic systems because chaos can be found in almost any system. A typical example is a quantum particle in a lattice which is subject to AC and DC fields Eq.~(\ref{ws_chaos}). This problem attracted much attention since the 1980's and was analyzed by using the single-band approximation in dozens of papers, without saying a word about chaos. However, the single-band approximation imposes severe restrictions on the system parameters and, if we enter the parameter region where the single-band approximation is not justified, we immediately face chaos. Moreover, in many systems we meet chaos even if we remain within the frame of single-band approximation. The bright example is the Bose-Hubbard model considered in Sec.~\ref{sec3}.

It was shown in  Sec.~\ref{sec3} that in bosonic systems  chaos is induced by a weak inter-particle interaction and, in this sense, it is a very common phenomenon.  The results of Sec.~\ref{sec3} indicate that usually chaos causes decoherence that leads, in particular, to the appearance of the dephasing operator Eq.~(\ref{decoherence}) in the master equation for the single-particle density matrix.  In the two-terminal transport problem, Sec.~\ref{sec4} and Sec.~\ref{sec5}, this dephasing is responsible for the crossover from the ballistic transport regime to the diffusive regime, where the particle current across the Bose-Hubbard chain is greatly suppressed.  Thus, in this example chaos plays a destructive role. However, the master equation with dephasing Lindblad operator is also the starting point for studying a bunch of phenomena known under the name  {\em environmentally assisted  transport} \cite{Rebe09,Kass12,Skal25,136}, where chaos plays a constructive role. 

In  Sec.~\ref{sec6} we paid considerable attention to a periodically driven chain of dissipative nonlinear oscillators, which is the classical counterpart of the driven dissipative Bose-Hubbard model. The key ingredients of this classical system are external driving, nonlinearity, and dissipation. It is known from the theory of dynamical systems that these ingredients, if they are taken together, are responsible for a variety of phenomena, including the phenomenon of {\em synchronization} \cite{Stro00}.  Some results on quantum synchronization in the driven dissipative  Bose-Hubbard system, which recover the ballistic transport for interacting bosons, are reported in Ref.~\cite{126}. There is no doubt that in the future we shall witness discoveries of new transport effects where chaos plays the key role.

\begin{ack}[Acknowledgments]

The Author acknowledges contributions of his co-authors in the original research papers,  results of which were used in the present tutorial review. The Author also apologizes to colleagues worldwide for not citing many relevant works due to the chapter length limit imposed by the publisher.
\end{ack}

\bibliographystyle{JHEP}%
\bibliography{kolovsky2}

\end{document}